\documentclass[11pt,draftcls,onecolumn]{IEEEtran}
\usepackage[latin1]{inputenc}
\usepackage{amsmath}
\usepackage{graphicx,color}
\usepackage{amssymb,url}

\newcommand{\QQ}{\mathbb{Q}}
\newcommand{\RR}{\mathbb{R}}
\newcommand{\ZZ}{\mathbb{Z}}
\newcommand{\NN}{\mathbb{N}}

\newcommand{\FF}{\mathbb{F}}

\newcommand{\cv}{\mathbf{c}}
\newcommand{\gv}{\mathbf{g}}
\newcommand{\mv}{\mathbf{m}}
\newcommand{\nv}{\mathbf{n}}
\newcommand{\rv}{\mathbf{r}}
\newcommand{\sv}{\mathbf{s}}
\newcommand{\tv}{\mathbf{t}}
\newcommand{\uv}{\mathbf{u}}
\newcommand{\vv}{\mathbf{v}}

\newcommand{\xv}{\mathbf{x}}
\newcommand{\yv}{\mathbf{y}}
\newcommand{\zv}{\mathbf{z}}

\newcommand{\Sc}{\mathcal{S}}

\newcommand{\Vc}{\mathcal{V}}

\newcommand{\vol}{{\rm vol}}
\newcommand{\Li}{{\rm Li}}

\newcommand{\yop}{y_{\mathsf{o.p.}}}

\newcommand{\SNR}{\mathsf{SNR}}

\newtheorem{example}{Example}
\newtheorem{defn}{Definition}
\newtheorem{prop}{Proposition}
\newtheorem{remark}{Remark}
\newtheorem{conject}{Conjecture}
\newtheorem{theo}{Theorem}

\begin{document}

\title{Lattice Codes for the Wiretap Gaussian Channel:
Construction and Analysis}

\author{Fr\'ed\'erique Oggier, Patrick Sol\'e and Jean-Claude Belfiore
\thanks{F. Oggier is with Division of Mathematical Sciences, School of
Physical and Mathematical Sciences, Nanyang Technological University,
Singapore. P. Sol\'e is with Telecom ParisTech, CNRS, UMR 5141, France and with Mathematics 
Department, King AbdulAziz Unversity, Jeddah, Saudi Arabia.
J.-C. Belfiore is with Telecom ParisTech, CNRS, UMR 5141, France.
Email:frederique@ntu.edu.sg,\{sole,belfiore\}@telecom-paristech.fr.
Part of this work appeared at ISITA 10~\cite{sec-gain} and at ITW 10~\cite{unimodular}.
}}
\maketitle
\begin{abstract}
We consider the Gaussian wiretap channel, where two legitimate players Alice
and Bob communicate over an additive white Gaussian noise (AWGN) channel, while Eve is eavesdropping, also through
an AWGN channel. We propose a coding strategy based on lattice coset encoding.
We analyze Eve's probability of decoding, from which we define the secrecy gain as
a design criterion for wiretap lattice codes, expressed in terms of the lattice theta series,
which characterizes Eve's confusion as a function of
the channel parameters. The secrecy gain is studied for even unimodular lattices,
and an asymptotic analysis shows that it grows exponentially in the dimension of
the lattice. Examples of wiretap lattice codes are given. \textcolor{black}{Interestingly, minimizing Eve's probability of error involves the same optimization of the theta series as does the flatness factor, another newly defined code design that characterizes lattice codes that achieve strong secrecy.}
\end{abstract}

%
%

\section{Introduction}

The wiretap channel was introduced by Wyner \cite{W-75} as a
discrete memoryless broadcast channel where the sender, Alice, transmits
confidential messages to a legitimate receiver Bob, in the presence of
an eavesdropper Eve. Wyner defined the perfect secrecy capacity as
the maximum amount of information that Alice can send to Bob while
insuring that Eve gets a negligible amount of information. He also
described a generic coding strategy known as coset coding,
used to encode together both data and random bits to confuse the eavesdropper.
The question of determining the secrecy capacity of many classes of
channels has been addressed extensively recently, yielding a plethora
of information theoretical results on secrecy capacity \cite{FnT-LPS}. 
\textcolor{black}{
In particular, the secrecy capacity of the Gaussian wiretap channel is known, 
and was established in \cite{Leung}.}

There is a sharp contrast with the situation of wiretap code designs,
where very little is known. Ozarow and Wyner proposed the so-called wire-tap
II codes \cite{OW-84} for a scenario where the channel to Bob is a noiseless
binary channel, while Eve experiences erasures. Recently polar wiretap codes
have been proposed for symmetric binary input channels \cite{MV-10,HS-10}.
The most exploited approach to get practical codes so far has been to use LDPC
codes, for binary erasure and symmetric channels (for example \cite{TDCMM-07}),
but also for Gaussian channels with binary inputs \cite{KHMBK-09}.

In this work, we consider lattice codes for Gaussian channels, where Alice uses
lattice coset encoding. Lattice codes for Gaussian channels have been considered
from an information theoretical point of view in \cite{HY-09} \textcolor{black}{in the setting of cooperative jamming}, \textcolor{black}{and more recently in \cite{laura,Ling}, 
where lattice codes have been considered for respectively the mod $\Lambda$ Gaussian wiretap channel, 
and the Gaussian wiretap channel. Both papers propose the so-called flatness factor as a new design criterion, and \cite{Ling} proves that nested lattice codes can achieve semantic and strong secrecy over the Gaussian wiretap channel.} 
We focus here on a code design criterion, which we derive from minimizing Eve's probability
of correctly decoding. 
\textcolor{black}{
More precisely, a wiretap lattice code consists of a pair of nested lattices $\Lambda_e \subset \Lambda_b$, where $\Lambda_b$ is a lattice designed to ensure reliability for Bob, while $\Lambda_e$ is a sublattice of $\Lambda_b$ that increases Eve's confusion. We show that Eve's probability $P_{c,e}$ of correctly decoding a message intended to Bob is bounded by a function that depends on the noise on Eve's channel, and on the theta series of the lattice $\Lambda_e$ at a particular point.
Interestingly, the theta series at that same point also provides an upper bound on the mutual information between Alice's message and Eve's received message \cite{Ling}.
Mimicking the way the coding gain quantifies how much reliability a particular coding strategy brings with respect to uncoded transmission, we define the secrecy gain to quantify how much confusion a specific lattice provides compared to using the $\ZZ^n$ lattice.}
An asymptotic study of the secrecy gain and wiretap lattice codes are further presented.

The paper is organized as follows. Section \ref{sec:model} recalls the channel model,
how lattice coset encoding is performed, while Section \ref{sec:proba} contains an analysis of Eve's probability
of correctly decoding, from which design criteria are deduced. The notion of secrecy
gain is defined, illustrated and interpreted in Section \ref{sec:The-secrecy-gain}.
It is further analyzed for even unimodular lattices in Section \ref{sec:unimodular}.
The asymptotic analysis which describes the behavior of the secrecy gain when the lattice dimension 
grows is presented in Section \ref{sec:Asymptotic-Analysis}. 
Finally some wiretap lattice codes can be found in Section \ref{sec:wiretap}.

%
%

\section{Coding Scheme for the Gaussian Wiretap Channel}
\label{sec:model}

\subsection{The Gaussian Wiretap Channel}

\begin{figure}[ht]
\noindent \begin{centering}
\includegraphics[width=8cm]{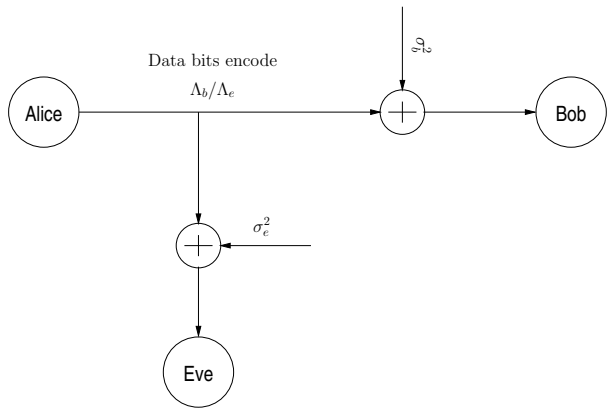}
\par\end{centering}
\caption{\label{fig:Wiretap-Gaussian}
The Gaussian wiretap channel between the sender Alice, and the two receivers
Bob and Eve.}
\end{figure}

We consider a Gaussian wiretap channel, that is, a broadcast channel where the
source (Alice) sends a signal to a legitimate receiver (Bob), while an
illegitimate eavesdropper (Eve) can listen to the transmission. It is modeled
by
\[
\begin{array}{ccl}
y &= & x + v_b\\
z &= & x + v_e,
\end{array}
\]
where $x$ is the transmitted signal, $v_b$ and $v_e$ denote the Gaussian
noise at Bob, respectively Eve's side, both with zero mean, and respective
variance $\sigma_b^2$ and $\sigma_e^2$ (see Figure \ref{fig:Wiretap-Gaussian}).
We assume that Alice knows Bob's channel, that is $\sigma_b$, as well as Eve's channel, $\sigma_e$, 
though we will also show how to handle the case where Eve's channel is unknown (see Section \ref{sec:wiretap}). 

Alice's encoder maps $l$ bits $s_1,\ldots,s_l$
from $\Sc=\{0,1\}$ to a codeword $\xv=(x_1,\ldots,x_n)\in\RR^n$, and over a
transmission of $n$ symbols, we get
\begin{equation}\label{eq:wiretap}
\begin{array}{ccl}
\yv &= & \xv + \vv_b\\
\zv &= & \xv + \vv_e.
\end{array}
\end{equation}

We consider the case where Alice uses lattice codes, namely
$\xv\in\Lambda_b$, where $\Lambda_b$ is an $n$-dimensional real lattice 
(we use the subscript $b$ to refer to the intended legitimate receiver Bob). 
She then encodes her $l$ bits 
into a point $\xv\in\Lambda_b$:
\[
\sv=(s_1,\ldots,s_l)\in \{0,1\}^l \mapsto \xv=(x_1,\ldots,x_n)\in \Lambda_b.
\]
Note that since Alice encodes a finite number $l$ of bits per codeword, she needs to choose a finite subset of $\Lambda_b$. The problem of finding a shaping region $\mathcal{R}$ is not addressed here.

We recall for the sake of completeness that a lattice $\Lambda$ is a discrete set of points
in $\RR^n$, which can be described in terms of its generator matrix $M$ by
\cite{OV-04,CS-98}
\[
\Lambda = \{ \xv = \uv M ~|~ \uv \in \ZZ^m \},
\]
\textcolor{black}{
where the $m$ rows of $M$ form a linearly independent set of vectors in $\RR^n$ (so that $m\leq n$)
which form a basis of the lattice. To every lattice $\Lambda$ is associated its dual lattice $\Lambda^\star$ defined 
as follows.
\begin{defn}\label{def:dual}
Let $\Lambda$ be a lattice with generator matrix $M$. We call its {\em dual lattice}
the lattice $\Lambda^\star$ with generator matrix $(M^{-1})^T$.
\end{defn}
}
For any lattice point $P_i$ of a lattice $\Lambda\subset \RR^n$,
its Voronoi cell is defined by
\[
\Vc_{\Lambda}(P_i)=\{\xv\in \RR^n,~d(\xv,P_i)\leq d(\xv,P_j)\mbox{ for all }P_j\in \Lambda\}.
\]
All Voronoi cells are the same,  thus $\Vc_{\Lambda}(P_i)=\Vc_{\Lambda}({\bf 0})=:\Vc\left({\Lambda}\right)$.
The volume of a lattice $\Lambda$ with generator matrix $M$ is by definition the volume $\vol(\Vc(\Lambda))$ of a Voronoi cell, that is
\[
\vol(\Vc(\Lambda))=\int_{\Vc(\Lambda)} d\xv =\det(MM^T)^{1/2}.
\]

\subsection{Wyner's Coset Encoding}

In order to confuse the eavesdropper, we use coset coding, as proposed
in \cite{W-75,OW-84}. The idea is that instead of having a one-to-one correspondence
between a vector of information bits and a lattice point, this vector of
information bits is mapped to a set of codewords, namely a coset, after
which the point to be actually transmitted is chosen randomly inside the
coset. Consequently, $k$ bits ($k\leq l$) of $\sv\in\{0,1\}^l$ will carry the information 
and $l-k$ bits, the randomness.  

More precisely, we partition the lattice $\Lambda_b$ into a union of
disjoint cosets of the form
\[
\Lambda_e+\cv,
\]
with $\Lambda_e$ a sublattice of $\Lambda_b$ and $\cv$ an $n$-dimensional
vector. We need $2^k$ cosets to be labeled by the information vector $\sv_d\in\{0,1\}^k$:
\[
\Lambda_b = \cup_{j=1}^{2^k} (\Lambda_e+\cv_j)
\]
which means that
\begin{equation}\label{eq:coset-card}
 \left|\Lambda_b/\Lambda_e\right|=2^k=\frac{\mbox{Vol}\left(\mathcal{V}\left(\Lambda_{e}\right)\right)}{\mbox{Vol}
\left(\mathcal{V}\left(\Lambda_{b}\right)\right)}.
\end{equation}

Once the mapping
\[
\sv_d \mapsto \Lambda_e+\cv_{j\left(\sv_d\right)}
\]
is done, Alice randomly chooses a point $\xv\in \Lambda_e+\cv_{j\left(\sv_d\right)}$ and
sends it over the wiretap channel.
This is equivalent to choose a random vector $\rv\in\Lambda_e$. The transmitted lattice point $\xv\in\Lambda_b$ is finally of the form
\begin{equation}\label{eq:xsent}
\xv = \rv + \cv \in \Lambda_e + \cv.
\end{equation}
We have denoted the sublattice $\Lambda_e$,
since it encodes the random bits that are there to increase Eve's confusion,
and is then the lattice intended for Eve.

The total rate $R$ is then
\[
R=R_s+R_e,
\]
where $R_s$ is the information bits rate intended to Bob, and $R_e$ is the random bit rate, all per (complex)
channel use:
\begin{equation}\label{eq:Rs}
R_s=\frac{2k}{n} \iff k=\frac{nR_s}{2},~R_e=\frac{2r}{n}\iff r=\frac{nR_e}{2},
\end{equation}
where $r$ is the number of random bits.

Intuitively, the meaning of this coding scheme is that we would like Eve to decode perfectly the lattice $\Lambda_e$ whose points are labeled 
by the random bits. This corresponds to the information-theoretic approach \cite{FnT-LPS} where it is shown that the secrecy capacity 
is equal to the difference between Bob's capacity and Eve's and thus, it is desirable that Eve's capacity is wasted in decoding 
random bits. 

\begin{example}\rm

Assume that the channel between Alice and Eve is corrupted
by an additive uniform noise. Even though this is not a realistic channel this perfectly 
illustrates the coset coding strategy. We will see that, in this case, it is enough to consider the $\ZZ$ lattice.  

\begin{figure}
\begin{center}
 
\includegraphics[width=7cm]{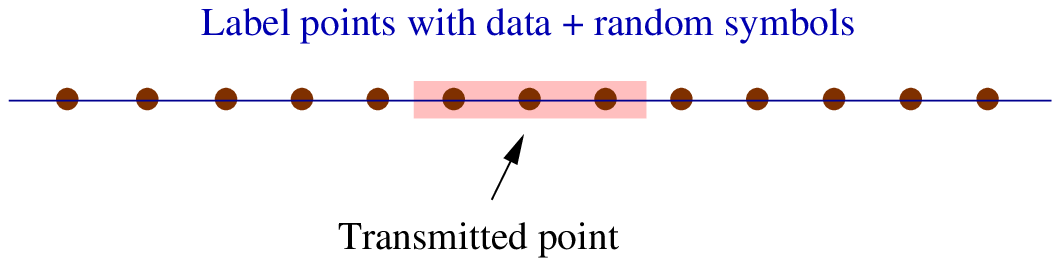}
\hspace{5mm}
\includegraphics[width=7cm]{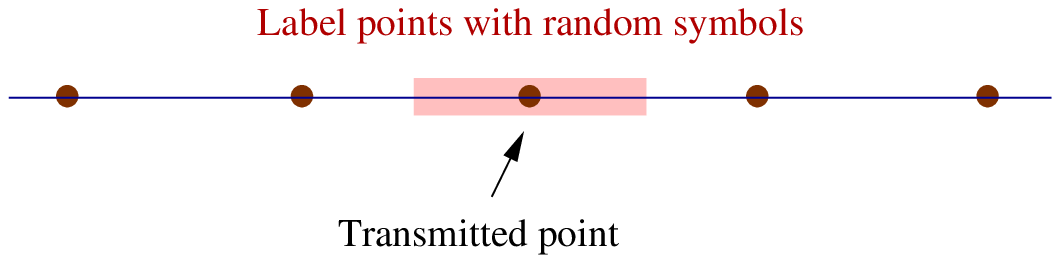}
\includegraphics[width=7cm]{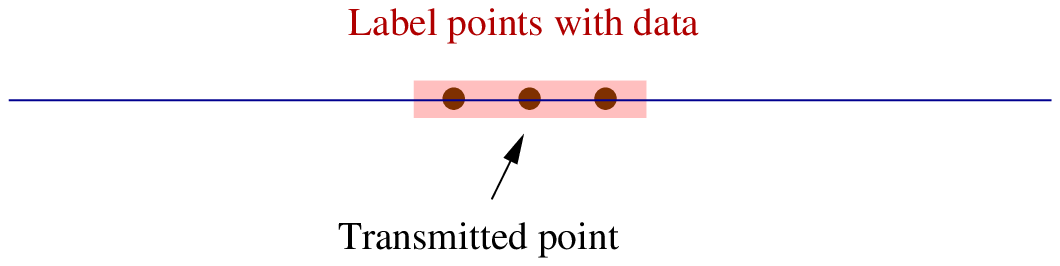}
\end{center}

\caption{
\label{fig:uniform}
Coset Coding with uniform noise: $\ZZ/m\ZZ$ with $m=3$.}
\end{figure}

Consider the one-dimensional case (see Figure \ref{fig:uniform}) where Alice sends one point $x\in\ZZ$. Eve receives 
\[
 y=x+v
\]
where $v$ is uniformly distributed over the interval $\left[ -\frac{m}{2},\frac{m}{2}\right ]$ for some $m\in \ZZ$, as 
shown on the upper left Figure \ref{fig:uniform}. 
To confuse Eve, Alice performs coset coding as follows: 
\begin{itemize}
 \item she performs the Euclidean division
\begin{equation}\label{eq:euc-div}
 x=mq+r, 0\leq r < m
\end{equation}

where the quotient $q$ carries the random symbols while the remainder $r$ carries the data.
\item she encodes random symbols using points in $m\mathbb{Z}$ (the quotient $q$)
while data symbols are mapped to elements of $\mathbb{Z}/m\mathbb{Z}$ (the remainder $r$). This is illustrated in 
the upper right and the lower parts of Figure \ref{fig:uniform}.
\end{itemize}

Now, as it can be seen in Figure \ref{fig:uniform}, Eve is able to detect with a zero-error probability the value of 
$q$ in Equation (\ref{eq:euc-div}) while all possible values of $r$ will be detected with probability $\frac{1}{m}$. This 
means that random symbols will be detected error-free when the confusion will be maximal for data symbols already 
when we use a one-dimensional lattice (that is $n=1$). 

Unfortunately, Gaussian noise is \emph{not} bounded: it \emph{requires} to use $n-$dimensional lattice codes. 
Table \ref{tab:multi-d} recalls the one-dimensional 
approach and shows the equivalent lattices with their respective cosets in the multi-dimensional approach 
required by the Gaussian channel. 

\begin{table}[ht]\label{tab:multi-d}
\noindent \begin{centering}
\begin{tabular}{|c|c|c|}
\cline{2-3} 
\multicolumn{1}{c|}{} & $1-$dimensional & $n-$dimensional\tabularnewline
\hline 
Transmitted lattice & $\mathbb{Z}$ & Fine lattice $\Lambda_{b}$\tabularnewline
\hline 
Random symbols & $m\mathbb{Z}\subset\mathbb{Z}$ & Coarse lattice $\Lambda_{e}\subset\Lambda_{b}$\tabularnewline
\hline 
Data & $\mathbb{Z}/m\mathbb{Z}$ & Cosets $\Lambda_{b}/\Lambda_{e}$\tabularnewline
\hline
\end{tabular}
\par\vspace{3mm}
\caption{From the example to the general scheme}
\end{centering}
\end{table}

%
%
\end{example}
\begin{example}\label{ex:example-2d} \rm

Consider the 2-dimensional lattice $2\ZZ^2$, that is
\[
2\ZZ^2=\{ (2x,2y),~x,y\in\ZZ\},
\]
and its cosets
\begin{eqnarray*}
2\ZZ^2+(0,1)&=&\{ (2x,2y+1),~x,y\in\ZZ\}, \\
2\ZZ^2+(1,1)&=&\{ (2x+1,2y+1),~x,y\in\ZZ\}, \\
2\ZZ^2+(1,0)&=&\{ (2x+1,2y),~x,y\in\ZZ\}.
\end{eqnarray*}
We note that if we take the union of $2\ZZ^2$ and its 3 cosets, we recover
the lattice $\ZZ^2$:
\[
\ZZ^2=\{(x,y),~x,y\in\ZZ\}=2\ZZ^2 \cup (2\ZZ^2+(0,1)) \cup (2\ZZ^2+(1,0))
\cup (2\ZZ^2+(1,1)).
\]
This is shown in Figure \ref{fig:Lattice--cosets}, where $2\ZZ^2$ is
represented by the triangles, $2\ZZ^2+(0,1)$ by the squares, $2\ZZ^2+(1,1)$
by the circles, and finally $2\ZZ^2+(1,0)$ by the stars.

\begin{figure}[ht]
\noindent \begin{centering}
\includegraphics[width=6cm]{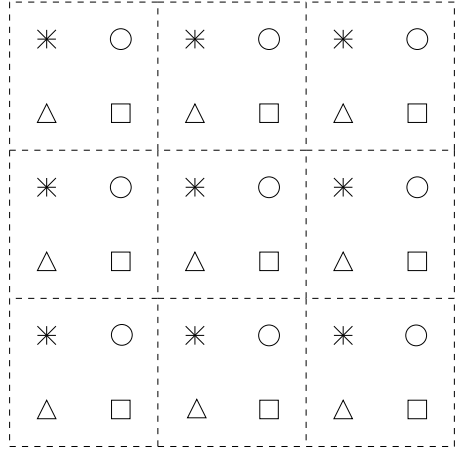}
\par\end{centering}
\caption{\label{fig:Lattice--cosets}The lattice $\mathbb{Z}^{2}$ seen as the
union of $4$ cosets.}
\end{figure}

Alice wants to communicate a message to Bob using the Gaussian wiretap channel
given in (\ref{eq:wiretap}). Assume that she can use $2$ bits per channel use,
she can then  label any of the above 4 cosets, say
\[
00 \mapsto 2\ZZ^2,~01 \mapsto (2\ZZ^2+(0,1)),~10 \mapsto (2\ZZ^2+(1,0)),
11 \mapsto (2\ZZ^2+(1,1)).
\]
To transmit the two bits $01$, she then randomly picks a point in the coset
$2\ZZ^2+(0,1)$, say $(2,3)$, and sends this point over the wiretap channel.

An interesting point to develop is the comparison, in terms of probability of correct decision for Eve, between
the scheme proposed here and 
the classical scheme using a $4-QAM$ constellation, that is, only using the symbols in the central square of Figure 
\ref{fig:Lattice--cosets}, to illustrate that coset coding does increase the confusion at the eavesdropper. 
For the classical $4-QAM$ constellation, the symbol probability of correct decision, at Eve's end, is given by \cite{proakis}
\begin{equation}\label{eq:pc-4qam}
 P_{c,e} = 1-2 Q\left( \sqrt{\frac{2E_b}{N_0}} \right)
\end{equation}
where $E_b$ is the energy per bit and $N_0=\sigma_e^2$ is the noise variance. $Q(x)$ is, as usual the error function defined as
\[
 Q(x)=\frac{1}{\sqrt{2 \pi}}\int_{x}^{+\infty}e^{-\frac{u^2}{2}}du.
\]

For the proposed scheme, the calculation of the probability of correct decision (for coset elements) at Eve's side can 
be done in the following way~: 
\begin{itemize}
 \item Decompose the $QAM$ constellation into its real and imaginary parts so that 
\[
P_{c,e}=\Pr\left\{ \hat{\xv}=\xv \right\}=\Pr\left\{ \hat{x}_1=x_1 \right\}\Pr\left\{ \hat{x}_2=x_2 \right\}
\]
where $\xv=x_1+ix_2$ is the transmitted $QAM$ symbol and $\hat{\xv}$ is the detected $QAM$ symbol. By symmetry of the 
constellation, we have 
\[
 \Pr\left\{ \hat{x}_1=x_1 \right\}=\Pr\left\{ \hat{x}_2=x_2 \right\}=:\Pr\left\{ \hat{x}=x \right\}.
\]

\item  Now, as can be seen on Figure \ref{fig:Lattice--cosets}
\begin{eqnarray*}
  \Pr\left\{ \hat{x}=x \right\}&=&\frac{1}{2}\left(\Pr\left\{ \hat{x}=\star,\triangle|x=\star,\triangle \right\}
+ \Pr\left\{ \hat{x}=\circ,\square|x=\circ,\square \right\} \right)\\
&=&\Pr\left\{ \hat{x}=\star,\triangle|x=\star,\triangle \right\}
=\Pr\left\{ \hat{x}=\circ,\square|x=\circ,\square \right\}
\end{eqnarray*}
so that 
\begin{equation}
 P_{c,e}=\left(\Pr\left\{ \hat{x}=\star,\triangle|x=\star,\triangle \right\}\right)^2.
\end{equation}
\item By summing over all coset representatives, we finally get that the probability of correct decision for Eve is
\begin{equation}\label{eq:pc-coset}
 P_{c,e} = \left[ 1-\frac{1}{3}\left( 5 Q\left( \sqrt{\theta} \right)-4 Q\left( 3\sqrt{\theta} \right)+3Q\left( 5\sqrt{\theta} \right)
-2Q\left( 7\sqrt{\theta} \right)+Q\left( 9\sqrt{\theta} \right) \right) \right]^2
\end{equation}
where $\theta=\frac{6}{35}\frac{E_b}{N_0}$. 

\end{itemize}

\begin{figure}[ht]\label{fig:pce}
\noindent \begin{centering}
\includegraphics[width=10cm]{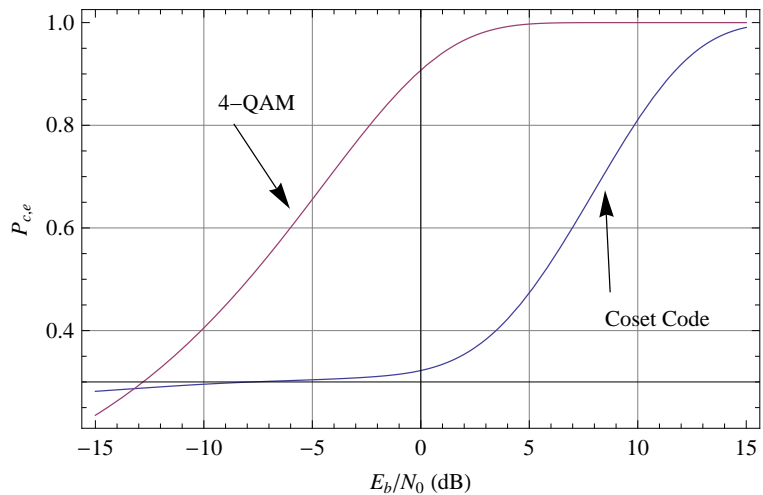}
\par\end{centering}
\caption{\label{fig:pc-curves}Probability of the eavesdropper correctly decoding the cosets. $4-QAM$ \textit{vs} coset scheme $\ZZ^2/2\ZZ^2$}
\end{figure}
As an illustration, $P_{c,e}$ as computed in (\ref{eq:pc-coset}) is plotted in Figure \ref{fig:pce} with the probability 
of correct decision for Eve when using a $4-QAM$ constellation. 
We observe that, if the $\SNR$ is either too big (above 15~dB) or too small (below -13~dB), there is no gain in using the 
coset scheme. Indeed, when the $\SNR$ goes below -13~dB, the size of the sphere of noise is such that it includes too many 
representatives of the correct coset, so that Eve's probability of guessing the coset that was sent is not negligible anymore. 
\end{example}

The example above shows the benefit of using coset encoding. However, it also illustrates that $P_{c,e}$ is 
no less than 0.3. We need to bring this threshold as low as possible (ideally tending to 0). This can be done by 
using multidimensional lattice coding in high dimension. 

\subsection{Lattice Coset Coding using Construction $A$}
There are several ways of getting lattice coset codes.
We will consider the so-called binary construction $A$ \cite{Ebeling} with binary codes.
Take the standard lattice $\ZZ^n \in \RR^n$ and reduce it modulo $2$~:
\[
\rho: \ZZ^n \rightarrow \left( \ZZ/2\ZZ \right)^n =\{0,1 \}^n.
\]
Let $C$ be a linear binary code with parameters $(n,\kappa,d)$, that is a map
from $\{0,1\}^\kappa$ to $\{0,1\}^n$ with minimum Hamming distance $d$. We can
partition an $n$-dimensional lattice $\Lambda$ as follows:
\[
\Lambda = 2\ZZ^n+C=\bigcup_{\cv_i\in C}(2\ZZ^n+\cv_i).
\]
This is also equivalent to say that $\Lambda$ is the preimage of $C$ in
$\ZZ^n$: $\Lambda=\rho^{-1}(C)$.

Example \ref{ex:example-2d} falls in this category.
Take the universe code $C$ with parameters $(2,2,1)$, given explicitly by
$\{(0,0),(0,1),(1,0),(1,1)\}$. Then
\[
\ZZ^2 = 2\ZZ^2+C=(2\ZZ^2+(0,0))\cup(2\ZZ^2+(0,1))\cup
(2\ZZ^2+(1,0))\cup (2\ZZ^2+(1,1)).
\]
Another 2-dimensional example is given by the checkerboard lattice $D_2$,
formed by integer vectors $(x_1,x_2)$ such that $x_1+x_2$ is even. Consider the
2-dimensional repetition code $\{(0,0),(1,1)\}$. Then
\[
D_2=2\ZZ^2+C=(2\ZZ^2+(0,0))\cup(2\ZZ^2+(1,1)).
\]
A more interesting example is the construction of the Sch\"affli
lattice $D_4$, formed by $(x_1,x_2,x_3,x_4)$ such that $x_1+x_2+x_3+x_4$ is even:
\[
D_4=2\ZZ^4+(4,3,2)
\]
where $(4,3,2)$ is the parity-check binary code of length $4$, dimension $3$ and minimum distance 2.

%
%
\section{Probability Analysis}\label{sec:proba}

\subsection{Coset Decoding}
After transmission over the Gaussian wiretap channel, Bob and Eve receive
respectively (see (\ref{eq:wiretap}) and (\ref{eq:xsent}))
\[
\begin{array}{ccll}
\yv &= & \xv + \vv_b &= \rv + \cv + \vv_b\\
\zv &= & \xv + \vv_e &= \rv + \cv + \vv_e,
\end{array}
\]
where we recall that $\rv\in\Lambda_e$ encodes the random bits, and
$\cv$ is the coset representative of minimum energy labeled by the
information bits.
Both Bob and Eve are interested in decoding the information bits, namely
in finding the correct coset that was sent. To do so, they need to find the
closest lattice point in $\Lambda_b$ to their respective received signal
$\yv$ or $\zv$, from which they deduce the coset to which it corresponds.

Now when transmitting a codeword $\xv$ in $\Lambda\subset\RR^n$ with Voronoi cell
$\Vc_{\Lambda}(\xv)$ over an additive white Gaussian noise channel with noise variance
$\sigma^2$, the decoder makes the correct decision if and only if the noisy
vector $\yv$ is in $\Vc_{\Lambda}(\xv)$, an event of probability
\[
\frac{1}{(\sigma\sqrt{2\pi})^n}\int_{\Vc_{\Lambda}(\xv)}e^{-||\yv-\xv||^2/2\sigma^2}d\yv.
\]
In our scenario, the probability $P_c$ of correct decision concerns not just
one point but a coset, and thus it is the probability that the received signal
lies in the union of the Voronoi regions of $\Lambda_b$, translated by points
of $\Lambda_e$.
Suppose that the lattice point $\xv=\rv+\cv\in\Lambda_b$ has been
transmitted, with $\rv\in\Lambda_e\cap\mathcal{R}\subset\Lambda_b$, where $\mathcal{R}$ is the shaping region of the constellation. 
The probability $P_c$ of finding 
the correct coset is thus,
\begin{equation}\label{eq:pc-general}
 P_c=\frac{1}{(\sigma\sqrt{2\pi})^n}
\sum_{\tv\in\Lambda_e\cap\mathcal{R}}\int_{\Vc_{\Lambda_b}(\xv+\tv)}e^{-||\yv-\xv||^2/2\sigma^2}d\yv.
\end{equation}

Since all terms in the sum of Equation (\ref{eq:pc-general}) are positive, we can upperbound 
it by extending the summation over the whole lattice $\Lambda_e$, which gives
\[
 P_c\leq\frac{1}{(\sigma\sqrt{2\pi})^n}
\sum_{\tv\in\Lambda_e}\int_{\Vc_{\Lambda_b}(\xv+\tv)}e^{-||\yv-\xv||^2/2\sigma^2}d\yv.
\]
If we take $M$ codewords from $\Lambda_b$, then and by doing the change of variable, 
$\uv=\yv-\xv-\tv$ we get
\begin{equation}\label{eq:Pe}
P_c\leq\frac{1}{(\sigma\sqrt{2\pi})^n}
\sum_{\tv\in\Lambda_e}\int_{\Vc\left(\Lambda_b\right)}e^{-||\uv+\tv||^2/2\sigma^2}d\uv.
\end{equation}

Accordingly, the probability $P_{c,b}$ of Bob's
(resp. $P_{c,e}$ of Eve's) correct decision is:
\begin{eqnarray}
P_{c,b}&\leq&\frac{1}{(\sqrt{2\pi}\sigma_b)^n}\sum_{\tv\in\Lambda_e}
\int_{\Vc\left(\Lambda_b\right)}e^{-\Vert \uv+\tv\Vert ^2/2\sigma_b^2}d\uv\label{eq:pcb-inter}\\
P_{c,e}&\leq&\frac{1}{(\sqrt{2\pi}\sigma_e)^n}\sum_{\tv\in\Lambda_e}
\int_{\Vc\left(\Lambda_b\right)}e^{-\Vert \uv+\tv\Vert ^2/2\sigma_e^2}d\uv.\label{eq:pce-inter}
\end{eqnarray}

Since Bob's received vector $\yv$ is most likely to lie in the Voronoi region of $\Lambda_b$ around the transmitted point \textcolor{black}{(Alice chooses $\Lambda_b$ to fit Bob's channel)},
the terms in $\tv$ different from ${\bf 0}$ in (\ref{eq:pcb-inter})
are negligible, which yields:
\begin{equation}\label{eq:Pcb}
P_{c,b}\leq \frac{1}{(\sqrt{2\pi}\sigma_b)^n}
\int_{\Vc(\Lambda_b)}e^{-\Vert \uv\Vert ^2/2\sigma_b^2}d\uv.
\end{equation}
This is now the familiar case of transmitting lattice points over the
Gaussian channel, for which it is known that $\Lambda_b$ should have a good
Hermite parameter, to get a good coding gain \cite{CS-98}.

\subsection{Eve's Probability of Correct Decision}

\textcolor{black}{
By (\ref{eq:pce-inter}), we need to evaluate 
\begin{equation}\label{eq:int-t}
\frac{1}{(\sqrt{2\pi}\sigma_e)^n}\sum_{\tv\in\Lambda_e}\int_{\Vc\left(\Lambda_b\right)}e^{-\Vert \uv+\tv\Vert ^2/2\sigma_e^2}d\uv = 
\int_{\Vc\left(\Lambda_b\right)}\frac{1}{(\sqrt{2\pi}\sigma_e)^n}\sum_{\tv\in\Lambda_e}e^{-\Vert \uv+\tv\Vert ^2/2\sigma_e^2}d\uv
\end{equation}
where $\tv\in\Lambda_e$. 
By denoting
\[
f(\tv)=e^{-\Vert \uv+\tv\Vert ^2/2\sigma_e^2},
\]
the Poisson formula for lattices (see (\ref{eq:poisslatt}) in the appendix) yields that
\[
\sum_{\tv\in\Lambda_e}f(\tv)=\vol(\Vc(\Lambda_e))^{-1}\sum_{\tv^\star\in\Lambda_e^\star}\hat{f}(\tv^\star)
\]
where $\Lambda^\star$ is the dual lattice of $\Lambda$ (see Definition~\ref{def:dual}).
We next compute $\hat{f}(\tv^\star)$, which by definition is
\begin{eqnarray*}
\hat{f}(\tv^\star)
&=& \int_{\RR^n} e^{-2\pi i \langle \tv^*, \vv \rangle} f(\vv)d\vv \\
&=& \int_{\RR^n} e^{-2\pi i \langle \tv^*, \vv \rangle} e^{\frac{-||\uv||^2-2\langle \uv,\vv \rangle - ||\vv||^2}{2\sigma_e^2}}d\vv\\
&=&
\prod_{j=1}^n e^{\frac{-u_j^2}{2\sigma_e^2}}\int_{\RR} e^{v_j\left(-2\pi i t_j^\star -\frac{2u_j}{2\sigma_e^2} \right)}e^{\frac{-v_j^2}{2\sigma_e^2}}dv_j\\
&=&
\prod_{j=1}^n \sqrt{2\pi\sigma_e^2}e^{\frac{-u_j^2}{2\sigma_e^2}}e^{2\sigma_e^2\left(\pi i t_j^\star +\frac{u_j}{2\sigma_e^2} \right)^2}
\end{eqnarray*}
using that
\begin{equation}\label{eq:int}
\int_\RR e^{-ax^2}e^{-2bx}dx=\sqrt{\pi/a}e^{b^2/a},~a>0.
\end{equation}
This yields 
\begin{eqnarray*}
\frac{1}{(\sqrt{2\pi}\sigma_e)^n}\sum_{\tv\in\Lambda_e}f(\tv)
&=&
\vol(\Vc(\Lambda_e))^{-1}\sum_{\tv^\star\in\Lambda^\star}
\prod_{j=1}^n e^{\frac{-u_j^2}{2\sigma_e^2}}e^{2\sigma_e^2\left(-\pi i t_j^\star +\frac{u_j}{2\sigma_e^2} \right)^2}\\
&=&
\vol(\Vc(\Lambda_e))^{-1}\sum_{\tv^\star\in\Lambda^\star}
e^{-\pi^22\sigma_e^2||\tv^*||^2}e^{-2\pi i \langle \tv^*,\uv \rangle}\\
&=&
\vol(\Vc(\Lambda_e))^{-1}\sum_{\tv^\star\in\Lambda^\star}
e^{-\pi^22\sigma_e^2||\tv^*||^2}\cos(2\pi \langle \tv^*,\uv \rangle)
\end{eqnarray*}
by noting that the sine term of the exponential averages out to zero when summing over all lattice points, and
\[
P_{c,e}\leq \vol(\Vc(\Lambda_e))^{-1}\int_{\Vc(\Lambda_b)}\sum_{\tv^\star\in\Lambda^\star}
e^{-\pi^22\sigma_e^2||\tv^*||^2}\cos(2\pi \langle \tv^*,\uv \rangle)d\uv.
\]
Now the cosine term takes it maximum value (that is 1) when $\uv \in \Lambda$, and we further get 
\begin{eqnarray*}
P_{c,e}
&\leq &\vol(\Vc(\Lambda_e))^{-1}\int_{\Vc(\Lambda_b)}\sum_{\tv^\star\in\Lambda^\star}
e^{-\pi^22\sigma_e^2||\tv^*||^2}d\uv \\
&=&
\frac{\vol(\Vc(\Lambda_b))}{\vol(\Vc(\Lambda_e))}
\sum_{\tv^\star\in\Lambda^\star}e^{-\pi^22\sigma_e^2||\tv^*||^2}.
\end{eqnarray*}
To obtain an expression which depends on $\Lambda$ instead of $\Lambda^\star$, we
denote this time
\[
f(\tv^\star)=e^{-2\pi^2\sigma_e^2\Vert \tv^\star \Vert ^2},
\]
and the Poisson formula for lattices (see (\ref{eq:poisslatt}) in the appendix) now gives that
\[
\sum_{\tv^\star \in\Lambda_e^\star}f(\tv^\star)=\vol(\Vc(\Lambda_e))\sum_{\tv\in\Lambda}\hat{f}(\tv)
\]
where $\hat{f}(\tv)$ is
\begin{eqnarray*}
\hat{f}(\tv)
&=& \int_{\RR^n} e^{-2\pi i \langle \tv, \vv \rangle} f(\vv)d\vv \\
&=& \int_{\RR^n} e^{-2\pi i \langle \tv, \vv \rangle} e^{- 2\pi^2\sigma_e^2||\vv||^2}d\vv\\
&=&
\prod_{j=1}^n \int_{\RR} e^{-2\pi i t_j v_j}e^{-2\pi^2\sigma_e^2v_j^2}dv_j\\
&=&
\left(\frac{1}{\sqrt{2\pi\sigma_e^2}}\right)^n \prod_{j=1}^n e^{\frac{-t_j^2}{2\sigma_e^2}}\\
&=&
\left(\frac{1}{\sqrt{2\pi\sigma_e^2}}\right)^n e^{-\frac{||\tv||^2}{2\sigma_e^2}}
\end{eqnarray*}
using that (\ref{eq:int}).
Finally the probability of making a correct decision for Eve is summarized by
\begin{equation}\label{eq:Pce}
P_{c,e}\leq\frac{1}{(\sqrt{2\pi}\sigma_e)^n}\vol(\Vc(\Lambda_b))
\sum_{\tv\in\Lambda_e}e^{-\Vert \tv\Vert ^2/2\sigma_e^2}.
\end{equation}
We can equivalently rewrite it in terms of generalized SNR (GSNR) $\gamma_{\Lambda_e}(\sigma_e)$ as
\begin{equation}\label{eq:Pcegsnr}
P_{c,e}\leq\frac{\vol(\Vc(\Lambda_e))}{(2\pi\sigma_e^2)^{n/2}} \frac{\vol(\Vc(\Lambda_b))}{\vol(\Vc(\Lambda_e))}
\sum_{\tv\in\Lambda_e}e^{-\Vert \tv\Vert ^2/2\sigma_e^2}
=
\gamma_{\Lambda_e}(\sigma_e)^{n/2} 2^{-nR_s/2}
\sum_{\tv\in\Lambda_e}e^{-\Vert \tv\Vert ^2/2\sigma_e^2}
\end{equation}
where
\begin{equation}\label{eq:gsnr}
\gamma_{\Lambda_e}(\sigma_e)=\frac{\vol(\Vc(\Lambda_e))^{2/n}}{2\pi\sigma_e^2}
\end{equation}
is the generalized signal-to-noise ratio (GSNR), and using (\ref{eq:coset-card}) and (\ref{eq:Rs}).
}

We know how to design good codes for Bob's channel, and have
his probability of making a correct decision arbitrarily close to 1.
Our aim is thus to minimize the probability $P_{c,e}$ of Eve making a correct
decision, while keeping $P_{c,b}$ unchanged.
This is equivalent to minimize (\ref{eq:Pce}), that is to find a lattice
$\Lambda_{b}$ which is as good as possible for the Gaussian channel
\cite{CS-98}, and which contains a sublattice $\Lambda_e$ such that
\begin{equation}
\label{eq:cd}
\boxed{
\begin{array}{c}
\mbox{minimize w.r. $\Lambda_e$}\sum_{\tv\in\Lambda_e}e^{-\Vert \tv\Vert ^2/2\sigma_e^2}\\
\mbox{under the constraint } \log_2\left| \Lambda_b/\Lambda_e \right|=k.
\end{array}
}
\end{equation}
The constraint on the cardinality of cosets (or rate) is equivalent
to set the fundamental volume of $\Lambda_e$ equal to a constant.

It is natural to start by approximating the sum of exponentials by its terms
of higher order, namely
\begin{eqnarray}
\sum_{\tv\in\Lambda_e}e^{-\Vert \tv\Vert ^2/2\sigma_e^2}
&\simeq& 1+\sum_{\tv\in\Lambda_e,||\tv||=d_{\min}(\Lambda_e)}
e^{-\Vert \tv\Vert ^2/2\sigma_e^2}\nonumber \\
& = & 1+ \tau(\Lambda_e)e^{-d_{\min}(\Lambda_e)^2/2\sigma_e^2},\label{eq:kissing-number}
\end{eqnarray}
where $\tau(\Lambda_e)$ is the kissing number of $\Lambda_e$ which counts
the number of vectors of length $d_{\min}(\Lambda_e)$.
Thus as a first criterion, we should maximize $d_{\min}(\Lambda_e)$ while
preserving the fundamental volume of $\Lambda_e$, which is equivalent to
require for $\Lambda_e$ to have a good Hermite parameter
\[
\gamma_H(\Lambda)=\frac{d_{\min}^2(\Lambda)}{\det(MM^T)^{1/n}}
\]
after which we should minimize its kissing number.
However we cannot be content with this approximation, and
have to obtain a more precise analysis as will be shown later on.

%
%
\section{The Secrecy Gain: a Design Criterion\label{sec:The-secrecy-gain}}

Let us get back to the code design criterion (\ref{eq:cd}) and rewrite it in
terms of the theta series of the lattice considered.
Recall that given a lattice $\Lambda \subset\RR^n$, its {\em theta series}
$\Theta_\Lambda$ is defined by \cite{CS-98}
\begin{equation}
\label{def:theta}
\Theta_{\Lambda}(z)=
\sum_{\xv\in\Lambda}q^{\left\Vert \xv\right\Vert ^{2}},
~q=e^{i\pi z},\mathrm{Im}(z)>0.
\end{equation}

\begin{example}
Let us compute the theta series of $\ZZ^n$:
\begin{eqnarray*}
\Theta_{\ZZ^n}(q) &=& \sum_{\xv\in\ZZ^n}q^{||\xv||^2}\\
                  &=& \sum_{(x_1,\ldots,x_n)\in\ZZ^n}q^{x_1^2+\ldots+x_n^2}\\
                  &=& \sum_{x_1\in\ZZ}q^{x_1^2}\cdots\sum_{x_n\in\ZZ}q^{x_n^2}\\
                  &=& \left(\sum_{n\in\ZZ}q^{n^2}\right)^n\\
                  &=& \Theta_{\ZZ}(q)^n.
\end{eqnarray*}
\end{example}

Exceptional lattices have theta series that can be expressed as functions
of the Jacobi theta functions $\vartheta_{i}(q)$, $q=e^{i\pi z}$,
$\mathrm{Im}(z)>0$, $i=2,3,4$, themselves defined by
\begin{eqnarray}
\vartheta_{2}(q) & =\sum_{n=-\infty}^{+\infty}q^{\left(n+\frac{1}{2}\right)^{2}},
\label{eq:theta2}\\
\vartheta_{3}(q) & =\sum_{n=-\infty}^{+\infty}q^{n^{2}},
\label{eq:theta3}\\
\vartheta_{4}(q) & =\sum_{n=-\infty}^{+\infty}\left(-1\right)^{n}q^{n^{2}}.
\label{eq:theta4}
\end{eqnarray}

A few examples of theta series of exceptional lattices \cite{CS-98} are given in Table \ref{tab:Theta-series}.
\begin{table}[ht]
\noindent \begin{centering}
\begin{tabular}{|c|c|}
\hline
Lattice $\Lambda$ & Theta series $\Theta_{\Lambda}$\tabularnewline
\hline
\hline
Cubic lattice $\mathbb{Z}^{n}$ & $\vartheta_{3}^{n}$\tabularnewline
\hline
Checkerboard lattice $D_{n}$ & $\frac{1}{2}\left(\vartheta_{3}^{n}+\vartheta_{4}^{n}\right)$\tabularnewline
\hline
Gosset lattice $E_{8}$ & $\frac{1}{2}\left(\vartheta_{2}^{8}+\vartheta_{3}^{8}+\vartheta_{4}^{8}\right)$\tabularnewline
\hline
Leech lattice $\Lambda_{24}$ &
$\frac{1}{8}\left(\vartheta_{2}^{8}+\vartheta_{3}^{8}+\vartheta_{4}^{8}\right)^3-
\frac{45}{16}\left(\vartheta_{2}\cdot\vartheta_{3}\cdot\vartheta_{4}\right)^8$\tabularnewline
\hline
\end{tabular}
\par\end{centering}
~
\caption{\label{tab:Theta-series}Theta series of some exceptional lattices}
\end{table}

From (\ref{eq:cd}), we need to minimize
\begin{eqnarray*}
\sum_{\tv\in\Lambda_e}e^{-\Vert \tv\Vert ^2/2\sigma_e^2}
&=&\sum_{\tv\in\Lambda_e} \left( e^{-1/2\sigma_e^2}\right)^{||\tv||^2} \\
&=&\sum_{\tv\in\Lambda_e} \left( (e^{i\pi})^{-1/2i\pi\sigma_e^2}\right)^{||\tv||^2} \\
&=& \Theta_{\Lambda_{e}}\left(z= \frac{-1}{2i\pi\sigma_{e}^{2}} \right)
\end{eqnarray*}
with $q=e^{i\pi z}$ and
\[
{\rm Im}\left(\frac{-1}{2i\pi\sigma_{e}^{2}}\right)=
{\rm Im}\left(\frac{i}{2\pi\sigma_e^2}\right)>0.
\]
Thus to minimize Eve's probability of correct decision is equivalent to
minimize $\Theta_{\Lambda_{e}}(z)$ in $z=i/2\pi\sigma_e^2$, under the constraint
that $\log_2\left| \Lambda_b/\Lambda_e \right|=k$. To approach this
problem, let us set $y=-iz$ and restrict to real positive values of $y$.
We are now interested in minimizing
\[
\Theta_{\Lambda_e}(y)=\sum_{\tv\in\Lambda_e}q^{\left\Vert \tv \right\Vert ^{2}},
~q=e^{-\pi y},y>0,
\]
over all possible $\Lambda_e$, in the particular value of $y$ corresponding to $z=i/2\pi\sigma_e^2$, namely
\begin{equation}\label{eq:y}
y=\frac{1}{2\pi\sigma_e^2}.
\end{equation}

\textcolor{black}{
\begin{remark}
From an information theory point of view, the information leaked to the eavesdropper is measured in terms 
of equivocation, that is $H(S^l|Z^n)$, where $S$ and $Z$ denote random variables corresponding respectively to the data and the message received by Eve. The best possible secrecy is achieved when $H(S^l|Z^n)=H(S^l)$, or equivalently when
\[
I(S^l;Z^n)=H(S^l)-H(S^l|Z^n)=0.
\]
How to design codes using the mutual information $I(S^l;Z^n)$ as a characterization of secrecy is not yet well understood. Recent progresses appeared in \cite[Theorem 5]{Ling}, where it was shown for the Gaussian wiretap channel that
\[
I(S^l;Z^n)\leq 8\epsilon_{n}nR-8\epsilon_{n}\log 8\epsilon_{n}=\epsilon_{n} (8nR-8\log 8\epsilon_{n}) ,
\]
where \cite[Proposition 1]{Ling}
\[
\epsilon_{n} =\gamma_{\Lambda_e}(\sigma_e)^{n/2}\Theta_{\Lambda_e}(1/2\pi\sigma_e^2)-1,
\]
and $\gamma_{\Lambda_e}(\sigma_e)$ is the
generalized signal-to-noise ratio defined in (\ref{eq:gsnr}). Both this information theory approach and our error probability approach agree on the fact that $\Theta_{\Lambda_e}(1/2\pi\sigma_e^2)$, that is the theta series of the lattice $\Lambda_e$ intended for Eve at the point $1/2\pi\sigma_e^2$ should be minimized. This bound is computed assuming a specific coding scheme, which takes into account a power constraint. Note that when we let the power grow, which corresponds to the scenario of the current paper, the way Alice encodes her message corresponds to choosing a point uniformly at random in a given coset, as is the case here. 
The interested reader may refer to \cite{Ling} for the connection between the flatness factor $\epsilon_{\Lambda_e(\sigma_e)}$ and the notion of strong secrecy.
\end{remark}
}

\subsection{Definition of Strong and Weak Secrecy Gains}

\textcolor{black}{
Mimicking the way the coding gain captures the benefit of a good coding strategy with respect to no coding in terms of probability of error, we introduce the {\em (strong) secrecy gain} to characterize how a good lattice $\Lambda_e$ increases the confusion at the eavesdropper, compared to choosing $\Lambda_e=\ZZ^n$.
}
\begin{defn}\label{def:secrecygain}
The \textit{strong secrecy gain} $\chi_{\Lambda,\mathrm{strong}}$ of an $n-$dimensional lattice
$\Lambda$ is defined by
\[
\chi_{\Lambda,\mathrm{strong}}= \sup_{y>0}\Xi_{\Lambda}(y), 
\]
where $\Xi_{\Lambda}(y)$ is the secrecy function of $\Lambda$, defined as follows.
\end{defn}
\begin{defn}\label{def:secrecyfunct}
Let $\Lambda$ be an $n-$dimensional lattice of volume $\lambda^n$. The \textit{secrecy function}
of $\Lambda$ is given by
\[
\Xi_{\Lambda}(y)=\frac{\Theta_{\lambda \mathbb{Z}^{n}}(y)}{\Theta_{\Lambda}(y)}
\]
defined for $y>0$.
\end{defn}

\begin{figure}[ht]
\begin{centering}
\includegraphics[width=10cm]{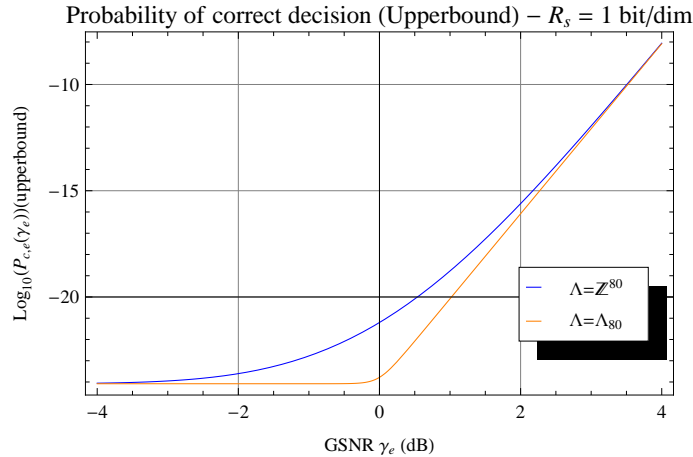}
\caption{
\label{fig:theta80}
A comparison between $\Theta_{\Lambda}{\gamma_e}_{\Lambda_e}$ for $\Lambda=\ZZ^{80}$ and 
$\Lambda^{80}$. $R_s=1$ bit per real dimension.}
\end{centering}
\end{figure}

\textcolor{black}{
These definitions deserve several observations.}
\begin{remark}
\begin{enumerate}
\item
The problem of minimizing $\Theta_{\Lambda_e}(y)$ under the rate
constraint $\log_2\left| \Lambda_b/\Lambda_e \right|=k$ means that the optimization
must be performed among lattices with the same volume. To do so, we fix as reference
a scaled version of the cubic lattice $\lambda \mathbb{Z}^{n}$, where $\lambda$ is a scaling factor which guarantees that $\Lambda_{e}$ and $\lambda \mathbb{Z}^{n}$
have the same fundamental volume, namely, $\lambda=\sqrt[n]{\vol(\Vc(\Lambda_e))}$.
\item
We are interested in the secrecy function 
at the chosen point $y=\frac{1}{2\pi\sigma_e^2}$. However, by considering $\sigma_e^2$ as a variable, 
and since we want to minimize the expression of Eve's probability of correct decision
in (\ref{eq:cd}),
it makes sense to further maximize the secrecy
function over $y>0$. 
\item
\textcolor{black}{
The secrecy function depends on $\sigma_e^2$. When Eve's channel is very noisy, there is no need for a subtle coding strategy ($\Lambda_e=\ZZ^n$ will do), and vice-versa, when Eve's channel is too good, wiretap coding cannot help ($\Lambda_e=\ZZ^n$ will again do). This is illustrated on Figure \ref{fig:theta80} where the behavior of the theta series of $\ZZ^{80}$ and of another lattice $\Lambda_{80}$ \footnote{See Subsection \ref{subsec:exteven} for more details about this lattice.}, both multiplied by the generalized SNR (GSNR), are compared, as a function of the GSNR (see (\ref{eq:gsnr})).
As a consequence, the secrecy function of a given lattice $\Lambda$ being the ratio of its theta series and the theta series of $\lambda \mathbb{Z}^{n}$ captures the region where wiretap coding is most meaningful, and provides an approximation of the ratio of the respective probabilities of correct decision.
}
\end{enumerate}
\end{remark}

Since the maximum value in Definition \ref{def:secrecygain} is not easy to calculate for a general lattice, we also introduce a weaker definition of secrecy gain. 
\textcolor{black}{
By (multiplicative) symmetry point, we mean a point $y_0$ such that
\[
\Xi_{\Lambda}(y_0\cdot y)=\Xi_{\Lambda}(y_0/y)
\]
for all $y>0$. We remark that the variable $y$ appears in the exponent,
explaining the multiplicative notation. One could alternatively express the
symmetry point in terms of $\log{y}$ and $\log{y_0}$, yielding
\[
\Xi_{\Lambda}({\log y_0}+\log{y})=\Xi_{\Lambda}(\log{y_0}-\log{y}).
\]
\begin{defn}\label{def:weak}
Suppose that $\Lambda$ is an $n$-dimensional lattice, whose secrecy function has a symmetry point $y_0$. 
Then the \textit{weak secrecy gain} $\chi_{\Lambda}$ of 
$\Lambda$ is given by
\[
\chi_{\Lambda}= \Xi_{\Lambda}\left( y_0 \right)=\frac{\Theta_{\lambda \mathbb{Z}^{n}}(y_0)}{\Theta_{\Lambda}(y_0)},
\]
\end{defn}
}
where we recall that $\lambda=\vol(\Vc(\Lambda))^{\frac{1}{n}}=|\det(M)|^{\frac{1}{n}}$.

\subsection{Lattices Equivalent to their Duals}

Let us consider the class of lattices $\Lambda$ such that $\Lambda$ is equivalent
to its dual $\Lambda^\star$, that is, the dual lattice $\Lambda^\star$ can be obtained from
the lattice $\Lambda$
by (possibly) a rotation, reflection, and change of scale $\alpha>0$:
\[
\Lambda \sim \alpha\Lambda^\star.
\]
In fact, if $\Lambda \sim \alpha\Lambda^\star$, then $\alpha$ cannot be
any positive number. Indeed, we deduce from the equivalence between both lattices
that
\[
\vol(\Vc(\Lambda))=\alpha^n \vol(\Vc(\Lambda^\star)).
\]
But since $\Lambda$ and  $\Lambda^\star$ are dual, then
\[
\vol(\Vc(\Lambda))=\frac{1}{\vol(\Vc(\Lambda^\star))}.
\]
From these two equalities, we get
\[
\alpha=\vol(\Vc(\Lambda))^{\frac{2}{n}}.
\]
If $\alpha=1$, we say that $\Lambda$ is \emph{isodual}. Alternatively
\begin{defn}\label{defn:isodual}
A lattice is {\em isodual} if it can be obtained from its dual by (possibly) a rotation or reflection.
\end{defn}

If $M$ is the generator matrix of $\Lambda$ and $(M^{-1})^T$ the one of its
dual, this means that $(M^{-1})^T=UMB$ where $U$ is a matrix with integer
entries and determinant $\pm 1$ and $B$ is a real orthogonal matrix. Thus the Gram matrix $G$ of $\Lambda$, which is by definition $G=MM^T$, is related to
the Gram matrix of its dual by $(M^{-1})^TM^{-1}=UMBB^TM^TU^T=UGU^T$.
A simple example of isodual lattice is $\ZZ^n$, since its generator matrix
$M={\bf I}_n$, and the one of its dual is $(M^{-1})^T={\bf I}_n$, and
both Gram matrices are the $n$-dimensional identity ${\bf I}_n$.
It follows from the definition of $\Lambda$ isodual that
$\Theta_{\Lambda}(y)=\Theta_{\Lambda^\star}(y)$, since the theta series depends
on the norm $||x||^2$, $x\in\Lambda$, which does not change by rotation or reflection of the lattice.
We are now ready to establish the weak secrecy gain of isodual lattices. 
\begin{prop}\label{prop:isodual}
 The secrecy function of an isodual lattice has a multiplicative symmetry point at $y=1$.
\end{prop}

\begin{IEEEproof}
 The secrecy function of an isodual lattice $\Lambda$
and the one of its dual $\Lambda^\star$ are the same:
\[
\Xi_{\Lambda}(y)=\frac{\Theta_{\mathbb{Z}^{n}}(y)}{\Theta_{\Lambda}(y)}=
\Xi_{\Lambda^\star}(y).
\]
Jacobi's formula (\ref{eq:jacobi}) gives, using that $\ZZ^n$ and $\Lambda$ are
isodual and have thus volume 1, that
\[
\begin{cases}
\Theta_{\ZZ^{n}}(y) & =y^{-\frac{n}{2}}\Theta_{\ZZ^{n}}\left(\frac{1}{y}\right)\\
\Theta_{\Lambda}(y) & =y^{-\frac{n}{2}}\Theta_{\Lambda^\star}\left(\frac{1}{y}\right)
\end{cases}
\]
and
\[
\Xi_{\Lambda}(y)=\frac{\Theta_{\ZZ^{n}}\left(\frac{1}{y}\right)}
{\Theta_{\Lambda^\star}\left(\frac{1}{y}\right)}
=
\Xi_{\Lambda}\left(\frac{1}{y} \right).
\]
This shows that $y_0=1$ is a multiplicative symmetry point for the secrecy function, which concludes the proof.  
\end{IEEEproof}
Consider again a lattice $\Lambda$ equivalent to its dual, though not necessarily isodual. The 
above result easily extends to this case. 

\textcolor{black}{
\begin{prop}\label{prop:equiv-dual}
The weak secrecy gain of a lattice equivalent to its dual is achieved at
\[
 y=\vol(\Vc(\Lambda))^{-\frac{2}{n}},
\]
that is
\[
\chi_{\Lambda}= \Xi_{\Lambda}\left( \frac{1}{\vol(\Vc(\Lambda))^{\frac{2}{n}}} \right).
\]
\end{prop}
}

\begin{IEEEproof}
We can in fact always scale the lattice $\Lambda$ as
\[
\Lambda'=\frac{1}{\vol(\Vc(\Lambda))^{\frac{1}{n}}}\Lambda
\]
so that $\Lambda'$ is isodual.
Now, since the theta series of a scaled lattice is
\[
\Theta_{\beta\Lambda}(y)=\Theta_{\Lambda}\left(\beta^2 y \right),
\]
with here $\beta=\vol(\Vc(\Lambda))^{-1}$, we deduce that
\[
\Xi_{\Lambda}\left(\vol(\Vc(\Lambda))^{-\frac{2}{n}}\cdot y\right)=
\Xi_{\Lambda'}(y)=\Xi_{\Lambda'}\left(\frac{1}{y} \right)=
\Xi_{\Lambda}\left(\frac{\vol(\Vc(\Lambda))^{-\frac{2}{n}}}{y} \right),
\]
which shows the existence, for $\Xi_{\Lambda}$, of a multiplicative symmetry point at
$y_0=\vol(\Vc(\Lambda))^{-\frac{2}{n}}$.
\end{IEEEproof}

\begin{conject}\label{conj:symmetry}
For a lattice equivalent to its dual, the weak secrecy gain and the strong secrecy gain coincide. In particular, 
this means that the secrecy function of isodual lattices achieves its maximum at $y=1$. 
\end{conject}
Note that a related problem has been addressed in \cite{C-06}: for a fixed dimension $n$, find the lattice
that minimizes $\Theta_{\Lambda}(y)$ for some value $y$. Unfortunately, the obtained results hold for values 
of $y$ belonging to a range which is not of interest. 

This conjecture is checked below for the lattices $E_8$ and $D_4$.

\subsection{Some Examples}
\begin{figure}[ht]
\begin{centering}
\includegraphics[width=8cm]{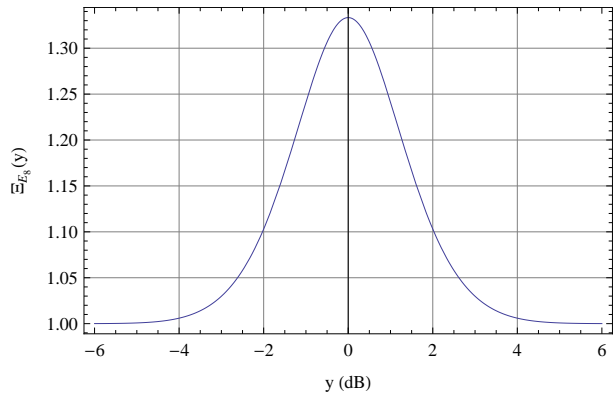}
\caption{
\label{fig:E8}
Secrecy function of $E_8$.}
\end{centering}
\end{figure}

\paragraph{The Gosset Lattice $E_8$}
The Gosset lattice is a famous $8$-dimensional lattice which can be described by
vectors of the form $(x_1,\ldots,x_8)$, $x_i\in\ZZ$, or $x_i\in\ZZ+1/2$, such that
$\sum x_i \equiv 0 \mod 2$. This lattice can be obtained by construction $A$ as
\[
\sqrt{2} E_8=2\ZZ^8+(8,4,4)
\]
where $(8,4,4)$ is the Reed-M\"uller code of length $8$ and dimension $4$, that is the extended binary Hamming $(7,4)$
code. $E_8$ is an isodual lattice and its theta series is given in Table
\ref{tab:Theta-series}. As it is isodual, the symmetry point of its secrecy function is $y_0=1$. Figure \ref{fig:E8} gives the
secrecy function of $E_8$. The symmetric point is also the point at which the secrecy function is maximized.
In all plots of the secrecy function, the horizontal axis will give $y$ in decibels ($10 \log_{10}(y)$) to enlighten
the symmetry point. Here, a multiplicative symmetry point equal to $1$ is, of course, represented by an additive
symmetry point equal to $0$ dB. We remark that the weak and the strong secrecy gains coincide.

\begin{figure}[ht]
\noindent \begin{centering}
\includegraphics[width=8cm]{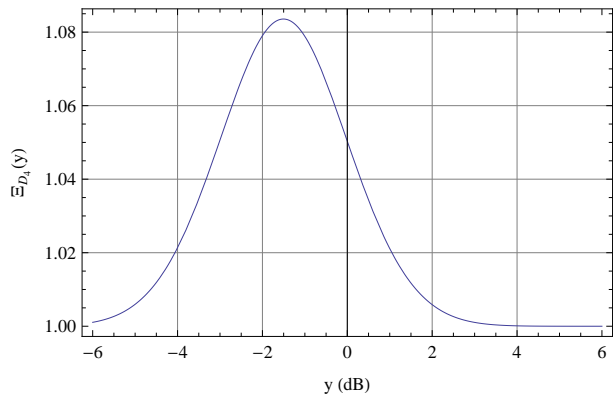}
\caption{
\label{fig:D4}
Secrecy function of $D_4$.}
\end{centering}
\end{figure}

\paragraph{The Sch\"affli lattice $D_4$}
$D_4$ is a $4-$dimensional lattice which is not isodual, but it is equivalent to its dual. Its fundamental
volume is $2$. This lattice can be obtained by construction $A$ as
\[
D_4=2\ZZ^4+(4,3,2)
\]
where $(4,3,2)$ is the binary parity-check code of length $4$.
The theta series of $D_4$ is also given in Table \ref{tab:Theta-series}. The multiplicative symmetry point is now
$y_0=\frac{1}{\sqrt{2}}$. Figure \ref{fig:D4} gives its secrecy function with a symmetry point equal to $-1.5$ dB
corresponding to $10 \log_{10} \left( \frac{1}{\sqrt{2}}\right)$. For this lattice also, the weak and the strong
secrecy gains again coincide.

\subsection{Operating Point of a Lattice\label{sec:Back-to-the}}

We are interested in how the secrecy gain is related
to the parameters of the Gaussian channel, through the proposed lattice
coset construction. \textcolor{black}{We restrict this discussion to lattices which are equivalent 
to their dual. In this case, from a system point of view, it is always possible to scale these lattices to normalize their volume to 1, in which case we obtain isodual lattices, which we showed have a symmetry point at $y=1$ (see Proposition \ref{prop:isodual}). }
Thanks to Conjecture \ref{conj:symmetry}, we will use the weak secrecy gain
instead of the strong one for isodual lattices and assume that we want the communication system to work at the value $y=1$.

In practice, this is obtained by scaling suitably the lattice $\Lambda_e$ for which we define correspondingly
its {\em operating point} $y_{\mathsf{o.p.}}$ as
\[
y_{\mathsf{o.p.}}=\mbox{Vol}\left(\mathcal{V}\left(\Lambda_{e}\right)\right)^{\frac{-2}{n}}.
\]

\begin{figure}[ht]
\noindent \begin{centering}
\includegraphics[width=10cm]{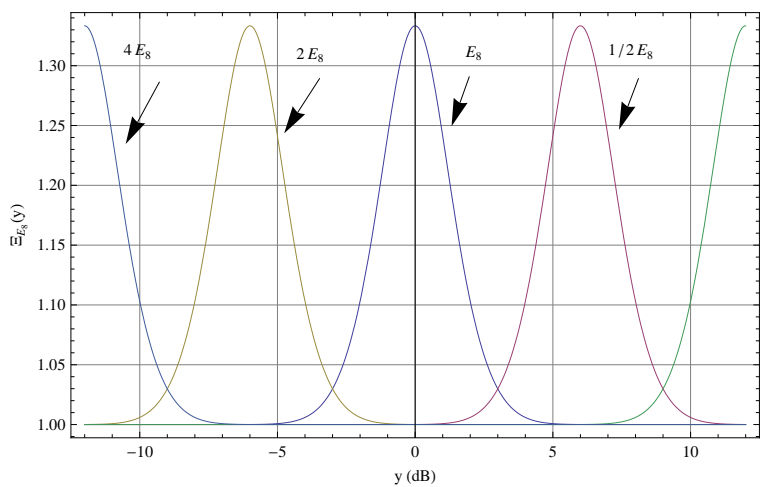}
\caption{
\label{fig:secgainE8}
Secrecy function for scaled versions of $E_8$.}
\end{centering}
\end{figure}

As an example, we see on Figure \ref{fig:secgainE8} how the operating point of
scaled versions of $E_8$ behaves with respect to the one of $E_8$.
For $2^mE_8,m\in\ZZ$, 
\[
 y_{\mathsf{o.p.}}=\mbox{Vol}\left(\mathcal{V}\left(2^mE_8\right)\right)^{\frac{-1}{4}}=
\left(2^{8m}\right)^{-\frac{1}{4}}=2^{-2m}
\]
that is $-6m$ dB. 

To fit the transmission rate, under the constraint (\ref{eq:coset-card}), that is
\[
 |\Lambda_b/\Lambda_e|=2^k=\frac{\mbox{Vol}\left(\mathcal{V}\left(\Lambda_e\right)\right)}{\mbox{Vol}
\left(\mathcal{V}\left(\Lambda_{b}\right)\right)},
\]
the fundamental volume of $\Lambda_{e}$ is scaled as
\[
\mbox{Vol}\left(\mathcal{V}\left(\Lambda_{e}\right)\right)=
2^{k}\mbox{Vol}\left(\mathcal{V}\left(\Lambda_{b}\right)\right)=
2^{\frac{nR_{b}}{2}}\mbox{Vol}\left(\mathcal{V}\left(\Lambda_{b}\right)\right).
\]
Thus
\begin{eqnarray*}
y_{\mathsf{o.p.}} & = & \mbox{Vol}\left(\mathcal{V}\left(\Lambda_{e}\right)\right)^{\frac{-2}{n}}\\
 & = & 2^{-R_{b}}\mbox{Vol}\left(\mathcal{V}\left(\Lambda_{b}\right)\right)^{\frac{-2}{n}}.
\end{eqnarray*}

Now, the average energy per complex symbol and per complex channel use if Alice sends a $Q-QAM$ constellation
with $Q=2^R$ points and minimum distance $2a$ is \cite{Forney}
\[
E_s(Q-QAM)=\frac{2(2^R-1)a^2}{3}.
\]
This can be easily extended to a $(Q-QAM)^\frac{n}{2}$ constellation, which can be seen as a 
cubically shaped subset of the $n$-dimensional lattice $2a\ZZ^n$:
\[
E_s(\left[2a\ZZ^n\right])=\mbox{Vol}\left(\mathcal{V}\left(2a\ZZ^n\right)\right)^{\frac{2}{n}}\frac{2^R-1}{6}
\]
where $[\Lambda]$ is a notation to refer to a cubically shaped subset of the lattice $\Lambda$. 
Now assuming that a finite constellation is carved from $\Lambda_b$ with a cubic shaping, its average energy
$E_s(\left[\Lambda_b\right])$ differs from the one of $\ZZ^n$ by its coding gain, which shows that we can approximate
the energy per complex channel use and per complex symbol of the signal sent by Alice by
\[
E_{s}(\left[\Lambda_b\right])\simeq  \mbox{Vol}\left(\mathcal{V}\left(\Lambda_b\right)\right)^{\frac{2}{n}}
\frac{2^R-1}{6}\simeq 2^{R}\mbox{Vol}\left(\mathcal{V}\left(\Lambda_{b}\right)\right)^{\frac{2}{n}}.
\]
Hence, we get
\[
y_{\mathsf{o.p.}}=2^{-R_s}E_{s}(\left[\Lambda_b\right])^{-1}2^{R},
\]
which with $\yop=\frac{1}{2\pi\sigma_e^2}$ from (\ref{eq:y}) gives
\[
\frac{1}{2\pi\sigma_e^2}=2^{-\left(R_s-R\right)}E_{s}(\left[\Lambda_b\right])^{-1}
\]
and finally
\begin{equation}
\boxed{
1=\frac{2^{-\left(R-R_{b}\right)}E_{s}(\left[\Lambda_b\right])}{2\pi\sigma_{e}^{2}}
=\frac{2^{-\left(R-R_{b}\right)}}{2\pi}\gamma_{e}
}
\label{eq:operating-point}
\end{equation}
where $\gamma_{e}=E_s/\sigma_e^2$ is Eve's signal to noise ratio.
This corresponds to a secrecy rate
\begin{equation}\label{eq:Rs-unimod}
R_s=R-\log_{2}\frac{\gamma_{e}}{2\pi}.
\end{equation}



%
%

\section{The Secrecy Gain of Unimodular Lattices}
\label{sec:unimodular}

Theta series are difficult to analyze in general, but nevertheless they have
nice properties for some families of lattices, such as even unimodular lattices,
which we will study in this section.

Let $\Lambda$ be a lattice with generator matrix $M$ and Gram matrix $G=MM^T$.
\begin{defn} \cite[Chap. 1]{Ebeling}
A lattice $\Lambda$ is {\em unimodular} if
\begin{enumerate}
\item $\Lambda$ is {\em integral}, i.e., its Gram matrix has entries in
$\mathbb{Z}$,
\item $\Lambda=\Lambda^{\star}$.
\end{enumerate}
It is furthermore {\em even unimodular} (or of {\em type II}) if
\[
\left\lVert \xv \right\rVert^2\equiv0 \mod 2, \forall x\in \Lambda.
\]
\end{defn}
Note that a unimodular lattice has fundamental volume equal to $1$.
Unimodular lattices are in particular isodual lattices, for which the weak
secrecy gain is reached in $y=1$, or $\log y=0$ (see Proposition \ref{prop:isodual}), and conjectured to be equal
to the strong secrecy gain.
We start by giving two examples of computations of the weak secrecy gain $\Xi(1)$ for two exceptional 
even unimodular lattices $E_8$ and $\Lambda_{24}$.


\subsection{The Secrecy Gain of Two Exceptional Unimodular Lattices}

The most important formulas we will use are related to Jacobi theta functions (\ref{eq:theta2})-(\ref{eq:theta4}) 
and can be found in \cite{jacobi-theta}.
They are
\begin{eqnarray}
\vartheta_{2}\left(e^{-\pi}\right) & = & \vartheta_{4}\left(e^{-\pi}\right)\nonumber \\
\vartheta_{3}\left(e^{-\pi}\right) & = & \sqrt[4]{2}\vartheta_{4}\left(e^{-\pi}\right)\label{eq:formulas}
\end{eqnarray}

\paragraph*{Gosset Lattice $E_{8}$}

We evaluate the value of the secrecy function $\Xi_{E_8}$ at the
point $y=1$ (Figure \ref{fig:E8} displays the secrecy function of $E_{8}$).
From Table \ref{tab:Theta-series}, we have that
\[
\Xi_{E_8}(y)=\frac{\vartheta_3(e^{-\pi})^8}
{\frac{1}{2}[\vartheta_2(e^{-\pi})^8+\vartheta_3(e^{-\pi})^8+\vartheta_4(e^{-\pi})^8]}.
\]
It is easier to look at $(\Xi_{E_8}(y))^{-1}$, which we evaluate in $y=1$:
\begin{eqnarray*}
\frac{1}{\Xi_{E_{8}}(1)}
& = & \frac{\tfrac{1}{2}\left(\vartheta_{2}(e^{-\pi})^{8}+\vartheta_{3}(e^{-\pi})^{8}+\vartheta_{4}(e^{-\pi})^{8}\right)}{\vartheta_{3}(e^{-\pi})^{8}}\\
&=&
\frac{1}{2}\left(1+\frac{2\vartheta_4(e^{-\pi})^8}{4\vartheta_4(e^{-\pi})^8}\right)\\
& = & \frac{3}{4}
\end{eqnarray*}
using (\ref{eq:formulas}).
We thus deduce that the secrecy gain of $E_{8}$ is
\[
\boxed{\chi_{E_{8}}=\Xi_{E_{8}}(1)=\frac{4}{3}=1.33333}
\]

\paragraph*{Leech Lattice $\Lambda_{24}$}

\begin{figure}
\noindent \begin{centering}
\includegraphics[width=8cm]{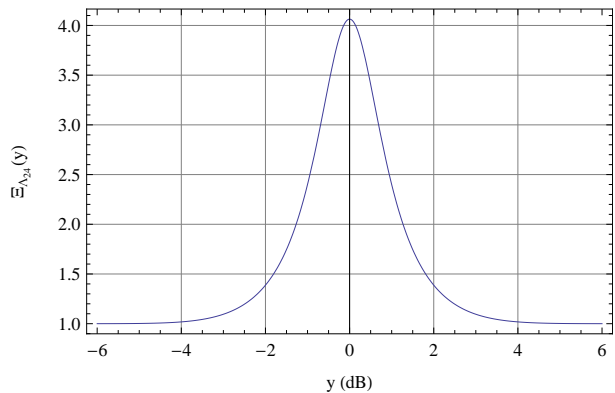}
\par\end{centering}
\caption{\label{fig:Secrecy-function-Leech}Secrecy function of $\Lambda_{24}$}
\end{figure}

From Table \ref{tab:Theta-series}, we get
\begin{eqnarray*}
\frac{1}{\Xi_{\Lambda_{24}}(1)}
& = & \frac{\frac{1}{8}\left(\vartheta_{2}(e^{-\pi})^{8}+\vartheta_{3}(e^{-\pi})^{8}
+\vartheta_{4}(e^{-\pi})^{8}\right)^{3}-\frac{45}{16}\vartheta_{2}(e^{-\pi})^{8}
\vartheta_{3}(e^{-\pi})^{8}\vartheta_{4}(e^{-\pi})^{8}}{\vartheta_{3}(e^{-\pi})^{24}}\\
 & =& \frac{\frac{6^3}{8}\vartheta_{4}(e^{-\pi})^{24}-\frac{45}{4}\vartheta_{4}(e^{-\pi})^{24}}{\vartheta_{3}(e^{-\pi})^{24}}\\
 & = & \frac{63}{256}
\end{eqnarray*}
again using (\ref{eq:formulas}), showing that the secrecy gain of $\Lambda_{24}$ is
\[
\boxed{\chi_{\Lambda_{24}}=\Xi_{\Lambda_{24}}(1)=\frac{256}{63}=4.0635}
\]
The secrecy function of $\Lambda_{24}$ is shown on Figure \ref{fig:Secrecy-function-Leech}.

\subsection{Theta series of Even Unimodular Lattices}
The theory of theta series of even unimodular lattices is well established\footnote{They are actually modular 
forms with integral weight \cite{Ebeling,CS-98}}. We first give some definitions
that will be useful for the calculation of the secrecy gain.
\begin{defn}
Consider the following two series
\begin{equation}\label{eq:Eisenstein}
E_{2k}(q)=1-\frac{4k}{B_{2k}}\sum_{m=1}^{+\infty}m^{2k-1}\frac{q^{2m}}{1-q^{2m}}
\end{equation}
and
\[
G_{2k}(q)=2\zeta(2k)E_{2k}(q),
\]
where $B_{k}$ are the Bernoulli numbers \cite{Serre-1} defined by
\begin{equation}\label{eq:Bk}
\frac{x}{e^{x}-1}=\sum_{l=0}^{\infty}B_{l}\frac{x^{l}}{l!},
\end{equation}
$k$ is an integer such that $k\geq 2$, $\zeta$ is the Riemann zeta function, and $q=e^{i\pi z}$,
$\mathrm{Im}(z)>0$. These series are referred to\footnote{The expression we use here as a 
definition is classically derived as a Fourier transform of another expression: $G_{2k}(\tau)=\frac{1}{2}\sum_{m,n}
\frac{1}{(m\tau+n)^{2k}}$, $(m,n)\neq 0$, $\mathrm{Im}(\tau)>0$.}
as Eisenstein series \cite[Chap. 2, \S 5]{Ebeling}.
\end{defn}

Note that these definitions hold for even $k'$, so that depending on the notations, one can write either $k'$ even, or as we choose here $k'=2k$, for $k$ a positive
integer. Furthermore, the argument $q$ can be either $q=e^{i\pi z}$ or
$q=e^{2i\pi z}$. Since so far we have always used $q=e^{i\pi}$, we keep this notation but then have to introduce a power of 2 in the exponent of $q$.

The Riemann zeta function $\zeta$ and the Bernoulli numbers are related by
\[
\zeta(2k)=(-1)^{k+1}\frac{(2\pi)^{2k}}{(2k)!}\left( \frac{B_{2k}}{2} \right),
\]
and it is known \cite{Serre-1} that $B_4=-1/30$, $B_6=1/42$. This allows us
to compute that
\begin{eqnarray*}
(60G_4(q))^3-27(140G_6(q))2
&=& ( 120 \zeta(4) )^3 E_4(q)^3-27(280 \zeta(6))^2 E_6(q)^2\\
&=& \frac{(2\pi)^{12}}{12^3}(E_4(q)^3-E_6(q)^2).
\end{eqnarray*}

We call
\begin{equation}\label{eq:delta}
\Delta(q)=\frac{1}{12^3}\left( E_4^3(q)-E_6^2(q)\right)
\end{equation}
the function that appears in the above computation, up to a factor of
$(2\pi)^{12}$ \cite[Chap. 2, \S 5]{Ebeling}, which is called the modular
discriminant\footnote{Different authors may or may not include the factor
$(2\pi)^{12}$ in the definition of modular discriminant.}.

Remarkably, theta series of all even unimodular lattices
can be expressed as polynomials in the two variables $E_4(q)$ and $\Delta(q)$:

\begin{prop}\label{theta-unimod}
If $\Lambda$ is an even unimodular lattice of dimension $n$, then
\begin{enumerate}
 \item $n=24m+8k$, for some positive integer $m$, and some $k\in\{0,1,2\}$ (as a consequence,  $n$ is a multiple of $8$),
 \item its theta series can be expressed, given $k,m$ in 1), as
\begin{equation}\label{eq:theta-unimod}
 \Theta_\Lambda(q)=E_4^{3m+k}(q)+\sum_{j=1}^m b_jE_4^{3(m-j)+k}(q)\Delta^j(q), \;b_j\in\QQ .
\end{equation}
\end{enumerate}
\end{prop}
\begin{IEEEproof}
 The proof can be found in \cite{Ebeling}.
\end{IEEEproof}

Now the two ``base'' series $E_4(q)$ and $\Delta(q)$ are simply related
to the Jacobi theta functions as \cite{CS-98}
\begin{equation}\label{eq:Eis-theta}
 \begin{cases}
E_{4}(q) & =\frac{1}{2}\left(\vartheta_{2}(q)^{8}+\vartheta_{3}(q)^{8}+\vartheta_{4}(q)^{8}\right)\\
\Delta(q) & =\frac{1}{256}\vartheta_{2}(q)^{8}\vartheta_{3}(q)^{8}\vartheta_{4}(q)^{8}
\end{cases}
\end{equation}

Equations (\ref{eq:Eis-theta}) and (\ref{eq:theta-unimod}) can be used to obtain a relation between the secrecy gain of an even unimodular lattice
$\Lambda$ of dimension $n=24m+8k$, on the one hand, and the ratios
\[
\rho_{E_4}=\frac{E_4\left( e^{-\pi}\right)}{\vartheta_3^8\left( e^{-\pi}\right)}
\]
and
\[
\rho_{\Delta}=\frac{\Delta\left( e^{-\pi}\right)}{\vartheta_3^{24}\left( e^{-\pi}\right)}
\]
on the other hand, since
\[
\frac{\Theta_{\Lambda}(q)}{\vartheta_3(q)^{24m+8k}}=
\left(\frac{E_4(q)}{\vartheta_3(q)^8}\right)^{3m+k}+
\sum_{j=1}^m b_j\left(\frac{E_4(q)}{\vartheta_3(q)^8}\right)^{3(m-j)+k}+
\left(\frac{\Delta(q)}{\vartheta_3(q)^{24}}\right)^j
\]
and thus
\begin{equation}\label{eq:chi-lambda}
 \frac{1}{\chi_{\Lambda}}=\rho_{E_4}^{3m+k}+\sum_{j=1}^m b_j\rho_{E_4}^{3(m-j)+k}\rho_{\Delta}^j, \;b_j\in\QQ .
\end{equation}

We can further deduce that
\begin{theo}
 The (weak) secrecy gain of an even unimodular lattice is a rational number.
\end{theo}
\begin{IEEEproof}
Note that
\[
\rho_{E_4}=\frac{1}{\chi_{E_8}}=\frac{3}{4},
\]
and using (\ref{eq:Eis-theta}) and (\ref{eq:formulas}), we get
\[
\rho_{\Delta}=\frac{\Delta\left( e^{-\pi}\right)}{\vartheta_3^{24}\left( e^{-\pi}\right)}=\frac{1}{2^{12}}.
\]
The proof then follows from Equation (\ref{eq:chi-lambda}).
\end{IEEEproof}

\subsection{Extremal Even Unimodular Lattices}
\label{subsec:exteven}

$E_{8}$ and $\Lambda_{24}$ are {\em extremal} even unimodular lattices
in dimensions $8$ and $24$ respectively \cite{CS-98}. We define below what is an extremal even unimodular lattice. 

Since we have that
\[
E_4(q)=1+\sum_{j=1}^\infty\alpha_jq^{2j}\mathrm{~and~}
\Delta(q)=\sum_{j=1}^\infty\beta_jq^{2j}
\]
for some coefficients $\alpha_j,\beta_j$, we have from (\ref{eq:theta-unimod})
that
\[
\Theta_{\Lambda}(q)=1+\sum_{j=1}^\infty\gamma_j q^{2j}
\]
for an even unimodular lattice. In order for it to be extremal, we set the coefficients
$\gamma_j=0$, $j=1,\ldots,m$, which yields a linear system of $m$ equations with $m$
unknowns given by $b_1,\ldots,b_m$. We then obtain the following development of
the theta series of $\Lambda$:
\[
\Theta_{\Lambda}(q)=1+\gamma_{2m+2}q^{2m+2}+O\left(q^{2m+4}\right)
\]
and consequently as upperbound for the minimum norm of $\Lambda$:
\begin{equation}\label{eq:upper-bound-dmin}
 \nu=\min_{x\in\Lambda\diagdown\{0\}}\left\lVert\xv\right\rVert^2\leq
2\lfloor m \rfloor+2.
\end{equation}
Unimodular lattices achieving the upperbound (\ref{eq:upper-bound-dmin}) are called
{\em extremal} and their theta series, determined by solving the above system of linear equations
in $b_j$, are called extremal theta series. They are given in Table
\ref{tab:Theta-series-extremal} for dimensions 8 to 80.
We notice that there is only one extremal theta series for a given dimension. Note that knowing the theta series does not give the corresponding lattice.
\begin{table}
\noindent \begin{centering}
\begin{tabular}{|c|c|c|}
\hline
Dimension & Lattice $\Lambda$ & $\Theta_{\Lambda}$\tabularnewline
\hline
\hline
$8$ & $E_{8}$ & $E_{4}$\tabularnewline
\hline
$24$ & $\Lambda_{24}$ & $E_{4}^{3}-720\Delta$\tabularnewline
\hline
$32$ & $BW_{32}$ & $E_{4}^{4}-960E_{4}\Delta$\tabularnewline
\hline
$48$ & $P_{48}$ & $E_{4}^{6}-1440E_{4}^{3}\Delta+125280\Delta^{2}$\tabularnewline
\hline
$72$ & $L_{72}$ & $E_{4}^{9}-2160E_{4}^{6}\Delta+965520E_{4}^{3}\Delta^{2}-27302400\Delta^{3}$\tabularnewline
\hline
$80$ & $L_{80}$ & $E_{4}^{10}-2400E_{4}^{7}\Delta+1360800E_{4}^{4}\Delta^{2}-103488000E_{4}\Delta^{3}$\tabularnewline
\hline
\end{tabular}
\par\end{centering}
\caption{\label{tab:Theta-series-extremal}Theta series of extremal lattices}
\end{table}

\begin{figure*}[t]
\noindent \begin{centering}
\includegraphics[width=0.4\textwidth]{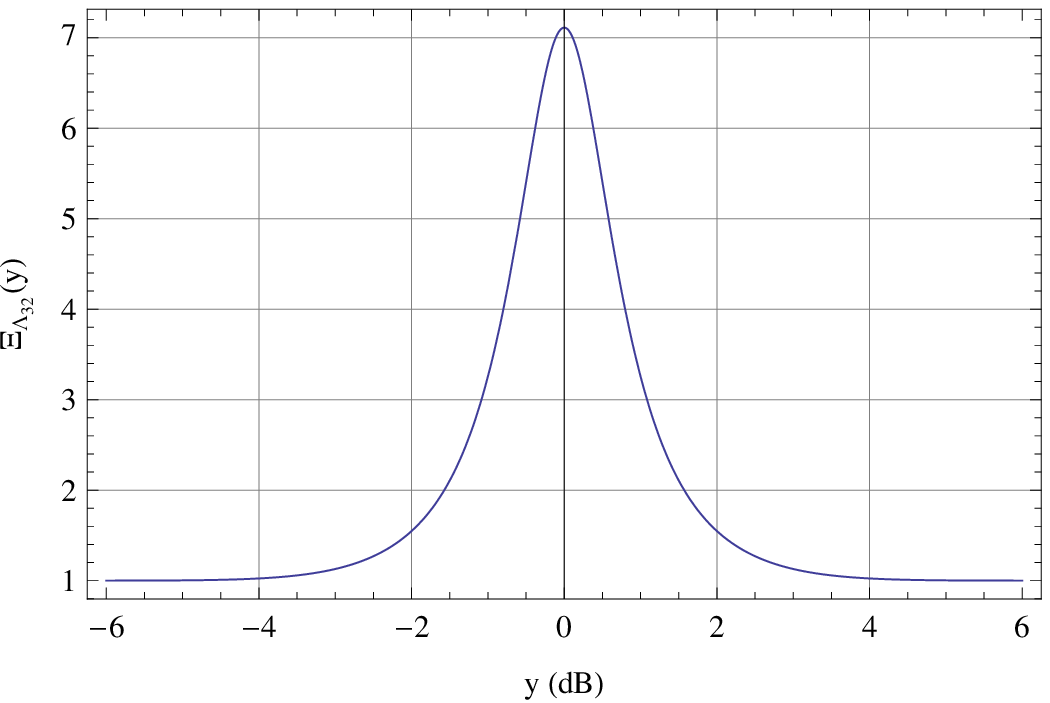}\quad{}\includegraphics[width=0.4\textwidth]{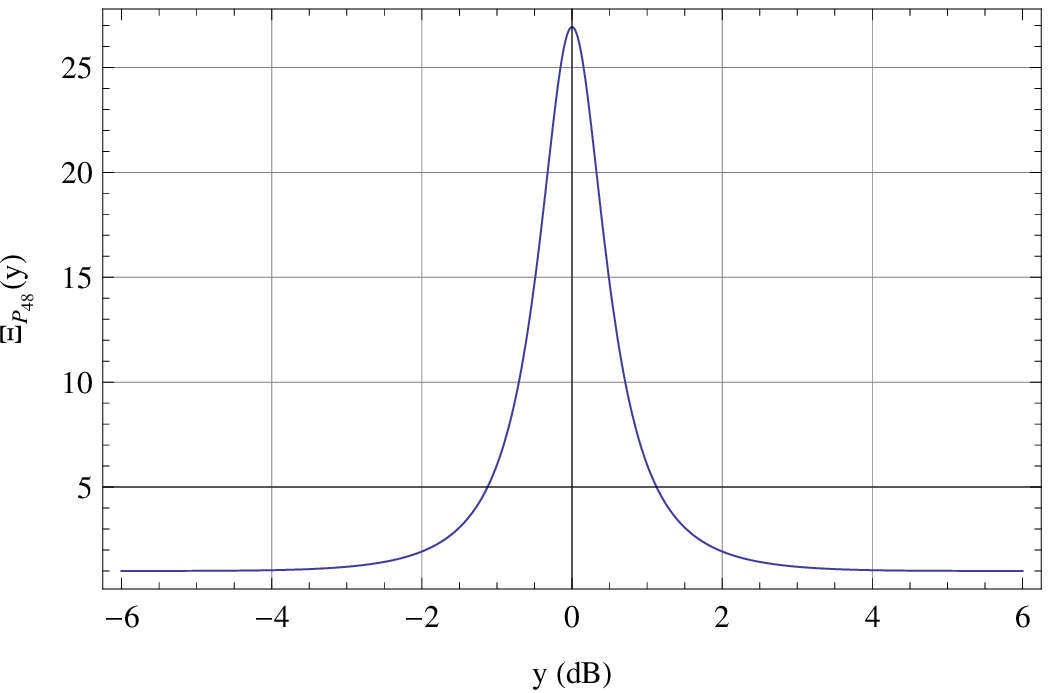}\bigskip{}
\par\end{centering}
\noindent \begin{centering}
\includegraphics[width=0.4\textwidth]{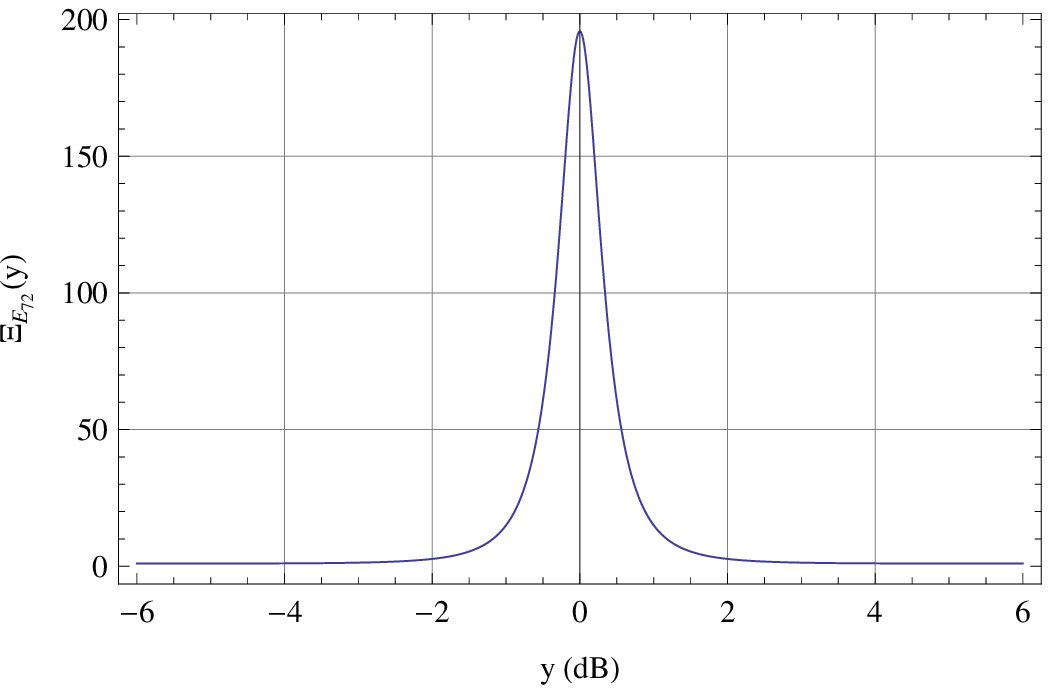}\quad{}\includegraphics[width=0.4\textwidth]{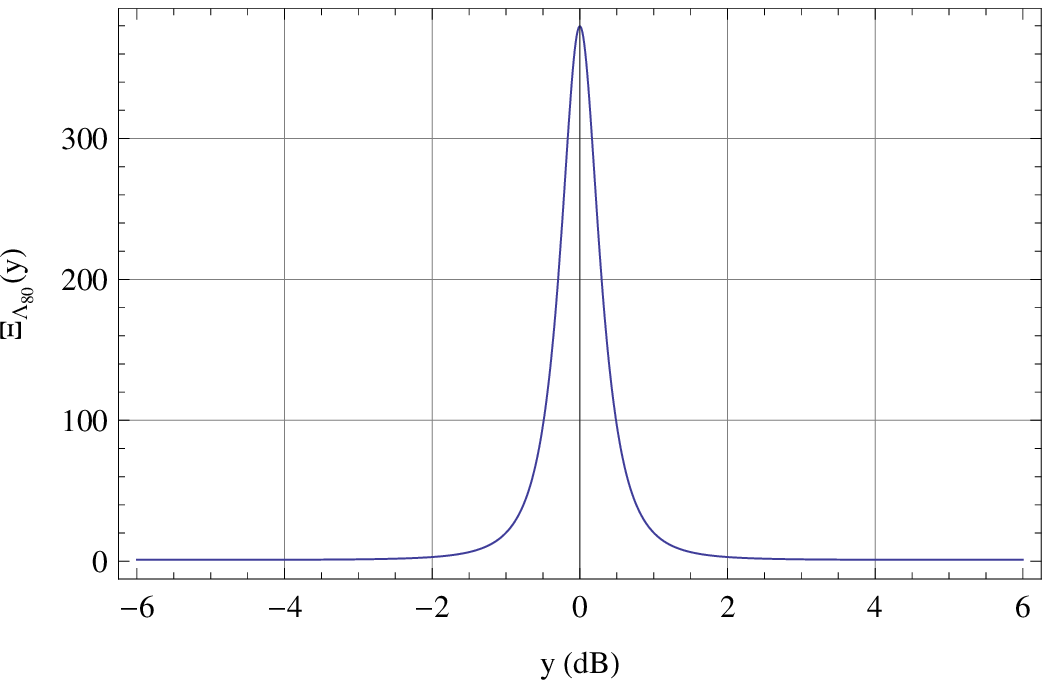}
\par\end{centering}
\caption{\label{fig:Secrecy-functions}Secrecy functions of extremal lattices
in dimensions 32, 48, 72 and 80}
\end{figure*}

We compute further values of secrecy gains for some extremal even unimodular
lattices in higher dimensions. The corresponding secrecy functions are shown on
Figure \ref{fig:Secrecy-functions}, while the different secrecy gains are summarized in
Table \ref{tab:Secrecy-gains-extremal}.

\subsubsection{Barnes-Wall lattice $BW_{32}$}

A $32$-dimensional extremal lattice is the Barnes-Wall lattice $BW_{32}$. Its theta series
is
\begin{eqnarray*}
\Theta_{BW_{32}}(q) & = & \frac{1}{16}\left(\vartheta_{2}(q)^{8}+\vartheta_{3}(q)^{8}+\vartheta_{4}(q)^{8}\right)\left[\left(\vartheta_{2}(q)^{8}+\vartheta_{3}(q)^{8}+\vartheta_{4}(q)^{8}\right)^{3}\right.\\
 &  & \left.-30\cdot\vartheta_{2}(q)^{8}\cdot\vartheta_{3}(q)^{8}\cdot\vartheta_{4}(q)^{8}\right]
\end{eqnarray*}
so that
\begin{align*}
\frac{1}{\Xi_{BW_{32}}(1)} & =\frac{1}{16}\left(1+\frac{1}{2}\right)\left[\left(1+\frac{1}{2}\right)^{3}-30\cdot\frac{1}{16}\right]\\
 & =\frac{9}{64},
\end{align*}
and finally its secrecy gain is
\[
\boxed{
\chi_{BW_{32}}=\frac{64}{9}\simeq7.11
}.
\]

\subsubsection{Lattice $P_{48p(q)}$}

There are two different extremal even unimodular lattices in dimension
$48$, $P_{48p}$ and $P_{48q}$ \cite[Chap. 5]{CS-98}, having, of
course the same theta series:
\begin{eqnarray*}
\Theta_{P_{48}}(q) & = & \frac{1}{2048}\left[3915\vartheta_{2}(q)^{16}\vartheta_{3}(q)^{16}\vartheta_{4}(q)^{16}\right.\\
 &  & -1440\vartheta_{2}(q)^{8}\vartheta_{3}(q)^{8}\vartheta_{4}(q)^{8}\left(\vartheta_{2}(q)^{8}+\vartheta_{3}(q)^{8}+\vartheta_{4}(q)^{8}\right)^{3}\\
 &  & \left.+32\left(\vartheta_{2}(q)^{8}+\vartheta_{3}(q)^{8}+\vartheta_{4}(q)^{8}\right)^{6}\right]\end{eqnarray*}
giving
\begin{align*}
\frac{1}{\Xi_{P_{48}}(1)} & =\frac{1}{2048}\left[\frac{3915}{256}-\frac{1440}{16}\left(1+\frac{1}{2}\right)^{3}+32\left(1+\frac{1}{2}\right)^{6}\right]\\
 & =\frac{19467}{524288}.
\end{align*}
Hence,
\[
\boxed{
\chi_{P_{48}}=\frac{524288}{19467}\simeq26.93
}.
\]

\subsubsection{Dimensions $72$ and $80$}

In the same way, from Table \ref{tab:Theta-series-extremal}, we can
compute the secrecy gain for an extremal unimodular even lattice in
dimension $72$ (found by G. Nebe \cite{Nebe}) and $80$. Note that two
examples of such lattices in dimension $80$ have been given in \cite{extremal-80}.
\begin{table}[ht]
\noindent \begin{centering}
\begin{tabular}{|c|c|c|c|c|c|c|}
\hline
Dimension & 8 & 24 & 32 & 48 & 72 & 80\tabularnewline
\hline
\hline
Secrecy gain & $1.3$ & $4.1$ & $7.11$ & $26.9$ & $195.7$ & $380$\tabularnewline
\hline
\end{tabular}
\par\end{centering}
\caption{\label{tab:Secrecy-gains-extremal}Secrecy gains of extremal lattices }
\end{table}
We have
\begin{eqnarray*}
\chi_{\Lambda_{72}} & = & \frac{134217728}{685881}\simeq195.69\\
\chi_{\Lambda_{80}} & = & \frac{536870912}{1414413}\simeq379.57
\end{eqnarray*}
We use the computation of the secrecy gain in dimension $80$ to illustrate two claims made earlier. 
\begin{enumerate}
 \item We saw, in Equation (\ref{eq:kissing-number}), the following approximation of the theta series: 
\[
 \sum_{\tv\in\Lambda_e}q^{-\Vert \tv\Vert ^2}
\approx 1+ \tau(\Lambda_e)q^{-d_{\min}(\Lambda_e)^2}.
\]
If we were to use this approximation to compute the secrecy gain, we would get 
\[
 \chi_{\Lambda_{80}}\approx \frac{1 + 160 e^{-\pi}}{1 + 1250172000 e^{-8\pi}}=7.7957
\]
instead of 379.57. This illustrates the importance of considering the whole theta series. 
\item 
Since the secrecy gain approximates the ratio of the respective probabilities of correct decision, we have that 
\[
 \frac{P_{c,e}\left(\ZZ^{80}\right)}{P_{c,e}\left(\Lambda_{80}\right)}
\approx \chi_{\Lambda_{80}}\approx 380. 
\]
We thus reduce Eve's probability of correct decision of a factor of $380$ by using $\Lambda_{80}$ instead of 
$\ZZ^{80}$. 
\end{enumerate}

%
%

\section{Asymptotic Analysis of the Secrecy Gain for Even Unimodular Lattices
\label{sec:Asymptotic-Analysis}}

In this section, we provide an asymptotic analysis of the secrecy gain $\chi_{\Lambda}$ for 
even unimodular lattices $\Lambda$\footnote{Theta series of even unimodular lattices are in fact modular forms for the whole modular group 
$SL_2\left(\ZZ\right)$, and all the results explained in this section actually rely on that property, 
though we are trying to use it as little as possible so as to make the paper accessible for people who are not familiar 
with the theory of modular forms.}. We first give a lower bound on the maximal value of $\chi_{\Lambda}$ over 
all even unimodular $n-$dimensional lattices, which only depends on the dimension $n$, after 
which we show more generally that as $n$ grows, 
the secrecy gain itself only depends on $n$, and not on the choice of a particular 
unimodular lattice.

\subsection{A lower bound on the maximal secrecy gain}

We propose here a lower bound on the theta series of unimodular lattices that maximizes
the secrecy gain, as a function of the dimension $n$. We then let $n$ grow to get an asymptotic bound. This result relies on the following Siegel-Weil formula for theta series
of even unimodular lattices.

\begin{theo} \cite{Serre-1}
Let $n\equiv0\:(\mbox{mod }8)$, $\Omega_{n}$ be the set of all inequivalent
even unimodular $n-$dimensional lattices and set $k=\frac{n}{2}$.
Then
\[
\sum_{\Lambda\in\Omega_{n}}\frac{\Theta_{\Lambda}(q)}{\left|\mathrm{Aut}(\Lambda)\right|}=M_{n}\cdot E_{k}\left(q\right)
\]
where \[
M_{n}=\sum_{\Lambda\in\Omega_{n}}\frac{1}{\left|\mathrm{Aut}(\Lambda)\right|},
\]
$E_{k}(q)$ is the Eisenstein series (\ref{eq:Eisenstein})\footnote{The index $k$ is an abuse of notation with 
respect to Definition (\ref{eq:Eisenstein}).}, 
and $\mathrm{Aut}(\Lambda)$
refers to the group of automorphisms of $\Lambda$.
\end{theo}

Let $\Theta_{\min}^{(n)}(e^{-\pi})=\min_{\Lambda\in\Omega_{n}}\Theta_{\Lambda}(e^{-\pi})$.
Then
\[
\Theta_{\min}^{(n)}(e^{-\pi})M_{n}\leq
\sum_{\Lambda\in\Omega_{n}}\frac{\Theta_{\Lambda}(e^{-\pi})}{\left|\mathrm{Aut}(\Lambda)\right|}=M_{n}E_{k}(e^{-\pi})
\]
so that
\[
\Theta_{\min}^{(n)}(e^{-\pi})\leq E_{k}(e^{-\pi}).
\]
Define
\[
\chi_{n}\triangleq\max_{\Lambda\in\Omega_{n}}\chi_{\Lambda}=
\frac{\vartheta_{3}^{n}\left(e^{-\pi}\right)}{\Theta_{\min}^{(n)}\left(e^{-\pi}\right)}
\]
where \cite{jacobi-theta}
\[
\vartheta_{3}\left(e^{-\pi}\right)=\frac{\pi^{\frac{1}{4}}}{\Gamma\left(\frac{3}{4}\right)}\simeq1.086
\]
to get
\[
\boxed{\chi_{n}\geq\frac{\vartheta_{3}^{n}\left(e^{-\pi}\right)}{E_{k}\left(e^{-\pi}\right)}=\frac{(1.086)^n}
{E_{k}\left(e^{-\pi}\right)}.
}
\]
Now
\[
E_{k}\left(e^{-\pi}\right)=1+\frac{2k}{\left|B_{k}\right|}\sum_{m=1}^{+\infty}\frac{m^{k-1}}{e^{2\pi m}-1}
\]
%
and for $k=4k'$ a multiple of $4$, we have
\begin{equation}
E_{4k'}\left(e^{-\pi}\right)
=1+\frac{8k'}{\left|B_{4k'}\right|}\sum_{m=1}^{+\infty}\frac{m^{4k'-1}}{e^{2\pi m}-1}\label{eq:E4k}. 
\end{equation}

An asymptotic expression of the Bernoulli number $\left |B_{4k'}\right |$ is
\begin{equation}\label{eq:Bernoulli-asymp}
 \left |B_{4k'}\right |=2\frac{(4 k')!}{(2 \pi )^{4k'}}.
\end{equation}

Now, as $e^{2\pi}\approx 535.5\gg 1$, we use 
\[
 e^{2\pi m}-1\sim e^{2\pi m},\; m\in \NN\backslash\{0\}
\]
 to get 
\[
 \sum_{m=1}^{+\infty}\frac{m^{4k'-1}}{e^{2\pi m}-1}\sim \sum_{m=1}^{+\infty}\frac{\left(e^{-2\pi}\right) ^m}{m^{1-4k'}}
=\Li_{1-4k'}\left(e^{-2\pi}\right)
\]
where $\Li_s(x)$ is the polylogarithm function defined as \cite{polylog}
\[
 \Li_s(x)=\sum_{m=1}^{+\infty}\frac{x^m}{m^s}.
\]
Now, we use the identity \cite{polylog}
\begin{equation}\label{eq:polylog}
 \Li_{1-4k'}\left(e^{-2\pi}\right)=\frac{(4k'-1)!}{(2\pi)^{4k'}}
\left[ \zeta(4k',1+i) + \zeta(4k',-i)\right]
\end{equation}

where $\zeta(s,x)$ is the so-called Hurwitz zeta function \cite{Hurwitz-zeta}. Combining the 3 equations below
\begin{eqnarray*}
 \Im\left(\zeta(4k',1+i)\right)&=&-\Im\left(\zeta(4k',-i)\right)\\
\lim_{k'\rightarrow +\infty}\Re\left(\zeta(4k',1+i)\right)&=&0\\
\lim_{k'\rightarrow +\infty}\Re\left(\zeta(4k',-i)\right)&=&1
\end{eqnarray*}
with Equation (\ref{eq:polylog}), we get
\begin{equation}\label{eq:Li-asymp}
 \lim_{k\rightarrow +\infty}\frac{\Li_{1-4k'}\left(e^{-2\pi}\right)}{(4k'-1)!/(2\pi)^{4k'}}=1. 
\end{equation}

Now we are ready to conclude. We combine Equations (\ref{eq:E4k}), (\ref{eq:Bernoulli-asymp}) and (\ref{eq:Li-asymp}) to obtain
\[
 \lim_{k'\rightarrow +\infty}E_{4k'}\left(e^{-\pi}\right)=1+
\frac{(4k'-1)!}{(2\pi)^{4k'}}\cdot\frac{(2\pi)^{4k'}}{(4k'-1)!}=2
\]

Since $n=4k'$, we finally conclude that
\begin{equation}
\boxed{\chi_{n}\gtrsim\frac{1.086^{n}}{2}}\label{eq:asymptotic-secrecy}
\end{equation}
which grows exponentially in $n$. %
\begin{figure}[ht]
\noindent \begin{centering}
\includegraphics[width=10cm]{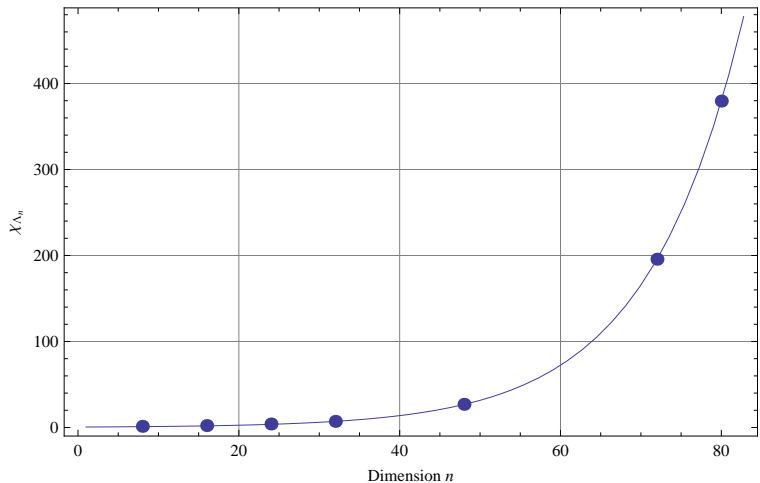}
\par\end{centering}
\caption{\label{fig:Secrecy-gain}Lower bound of the minimal secrecy gain as
a function of $n$ from Siegel-Weil formula. Points correspond to
extremal lattices.}
\end{figure}

Figure \ref{fig:Secrecy-gain} gives the asymptotic expression of
the secrecy gain as a function of the dimension $n$, as well as points
corresponding to extremal lattices in dimensions $8,16,24,32,48,72$
and $80$. 

This proves that there exists a family of even unimodular lattices whose
secrecy gains exponentially grows up with the dimension, which means
that Eve's probability of correct decision exponentially tends to
$0$. But as we can remark in Figure \ref{fig:Secrecy-functions},
around its maximum, the secrecy function becomes sharper and sharper
when $n$ grows, meaning that for high dimensions, the communication
system absolutely has to operate at the operating point ($y=1$ for unimodular
lattices).

\subsection{Behavior of the secrecy gain when $n$ grows}

Let us now look at the behavior of the secrecy gain when $n$ grows, which depends on 
the theta series of the corresponding even unimodular lattice $\Lambda$ of dimension $n$, a multiple of $8$. 
The main result used here is that the theta series of $\Lambda$ is given by 
\cite[Chap. 11]{Iwaniec}
\begin{equation}\label{eq:theta-decomp}
\Theta_{\Lambda}(q)=E_{k}(q)+S_{k}\left(q,\Lambda\right)
\end{equation}
where $E_{k}(q)$ is the Eisenstein series given in (\ref{eq:Eisenstein}) with $k=n/2$
and $S_{k,\Lambda}(q)$ is a function (a so-called cusp form) whose Fourier decomposition is of the form 
\[
S_{k}\left(z,\Lambda\right)=\sum_{m=0}^{\infty}a\left(m,\Lambda\right)e^{2i\pi mz}
\]
where the Fourier coefficients behave as \cite[Chap. 1]{Sarnak}
\[
a\left(m,\Lambda\right)=O_{\epsilon}\left(m^{\frac{k}{2}-\frac{1}{4}+\epsilon}\right).
\]
On the other hand, the Fourier decomposition of the Eisenstein series is
\[
E_{k}(z)=1+\frac{\left(2\pi\right)^{k}}{\zeta(k)\Gamma(k)}\sum_{m=1}^{\infty}\sigma_{k-1}(m)e^{2i\pi mz}
\]
where $\sigma_{k-1}(m)=\sum_{d|m}d^{k-1}$ is the divisor function which behaves 
as 
\[
\sigma_{k-1}(m)=O\left(m^{k-1}\right).
\]
By combining both Fourier coefficient estimations, we obtain that the Fourier coefficients of the theta series 
$\Theta_{\Lambda}(q)$ in (\ref{eq:theta-decomp}), when $n$ becomes large enough, is dominated by the Eisenstein 
series which only depends on the dimension $n$.
Consequently, when $n$ grows, the theta series of all even unimodular lattices behave like the Eisenstein series $E_k(q)$, 
which, in terms of secrecy gain $\chi_{\Lambda}$, means 
\[
 \chi_{\Lambda}\approx\frac{(1.08)^n}{2}
\]
for any $n-$dimensional ($n$ large enough) even unimodular lattice $\Lambda$. 

%
%
\section{Wiretap Lattice Codes}
\label{sec:wiretap}

We conclude this paper by giving some examples of code construction.

\subsection{An 8-dimensional $2$-level nested lattice code construction
\label{sub:2-level-nested-lattice}}

Suppose that Alice communicates with Bob using an $8$-dimensional lattice.
She needs to decide both $\Lambda_{b}$, that encodes bits for Bob, and $\Lambda_{e}$, 
that contains random bits intended for Eve. She can take $\Lambda_{b}=E_{8}$, since
this lattice has the best coding gain (Hermite constant) in
dimension $8$ \cite{CS-98}. Based on her knowledge of Bob's $\SNR$, $\gamma_b=E_s/\sigma_b^2$, 
and Bob's desired error probability, Alice decides the shaping region $\mathcal{R}$ and thus, 
the total rate $R=R_e+R_s$ of transmission. 

Now $\Lambda_{e}$ has to be a sublattice of $E_{8}$,
which first optimizes the secrecy gain. 
Since $E_8$ is an extremal lattice, all its scaled versions reach the lower bound on the 
maximal secrecy gain $\chi_8$ and consequently, we pick $\Lambda_e=2^mE_8$. Note that 
the scaling factor has to be a power of 2 since $\Lambda_e$ has to be a sublattice of 
$\Lambda_b=E_8$. This further quantizes the rate $R_s$ as follows:
\[
 \left|E_8/2^mE_8\right|=2^{8m}\Rightarrow R_s=\frac{2k}{8}=2m,
\]
and we have from (\ref{eq:Rs-unimod}) that
\begin{eqnarray*}
R-R_s &=& R_e \\
&=& \log_2(\gamma_e)-\log_2(2\pi)\\
&=& \frac{\gamma_e(\mathrm{dB})}{10}\log_{2}10-\log_2(2\pi)\\
&\simeq & \frac{\gamma_e(\mathrm{dB})}{10} (3.32)-(2.65).
\end{eqnarray*}
Thus since we are under the assumption that Alice knows Eve's $\SNR$, $\gamma_e$, she accordingly decides how many random bits to send.
For example
\begin{eqnarray*}
\gamma_e(\mathrm{dB})= 10 \mbox{ dB},~R_e \simeq 0.67 \\
\gamma_e(\mathrm{dB})= 20 \mbox{ dB},~R_e \simeq 4.
\end{eqnarray*}
Of course, the better Eve's $\SNR$, the more random bits are needed.
Now, $R$ is fixed by Bob's $\SNR$ while $R_e$ is given by Eve's $\SNR$ which constraints the data rate to be
\[
 R_s=R-R_e,~ R_s=2m,~m\in\ZZ.
\]
 
For example, if $R\approx6$ bits and Eve has a $\SNR$ of $\gamma_e=20$~dB, then
Alice can send $R_s=2$ bits per complex channel use, which means that $\Lambda_e=2E_8$.

The encoding is done via construction $A$, as explained in Section
\ref{sec:model}. First as already seen earlier,
\begin{equation}\label{eq:e82}
E_{8}=\sqrt{2}\mathbb{Z}^{8}+\frac{1}{\sqrt{2}}(8,4,4)
\end{equation}
where $C=(8,4,4)$ stands for the Reed-M\"uller code of length $8$ and dimension
$4$ and since $\ZZ^8= 2 \ZZ^8 + (8,8,1)$, we have
\begin{equation}\label{eq:e8}
E_8=\sqrt{2}\ZZ^8+\frac{1}{\sqrt{2}}(8,8,1)+\frac{1}{\sqrt{2}}(8,4,4).
\end{equation}
We denote by $C^\dagger$ the quotient code $C^\dagger=\FF_2^8/C$, or
equivalently
\[
\FF_2^8 = C+C^\dagger\mbox{ i.e., }(8,8,1)=(8,4,4)+C^\dagger,
\]
so that (\ref{eq:e8}) becomes
\[
E_8=\sqrt{2}\ZZ^8+\frac{1}{\sqrt{2}}(8,4,4)+\frac{1}{\sqrt{2}}C^\dagger+\frac{1}{\sqrt{2}}(8,4,4)
=\sqrt{2}\ZZ^8+\frac{1}{\sqrt{2}}C^\dagger,
\]
and
\[
\sqrt{2}\ZZ^8 =  E_8 + \frac{1}{\sqrt{2}}C^{\dagger} \Rightarrow 2\sqrt{2}\ZZ^8 =  2E_8 + \sqrt{2}C^{\dagger}.
\]
Combining with $E_{8}=\sqrt{2}\mathbb{Z}^{8}+\frac{1}{\sqrt{2}}(8,4,4)$, we finally obtain
a construction of $E_{8}$ using $2E_{8}$:
\[
E_{8}=2E_{8}+\frac{1}{\sqrt{2}}\left((8,4,4)+2C^\dagger\right).
\]
Now the $k=8$ bits of information are used to encode $(8,4,4)+2C^\dagger$
(4 bits for $(8,4,4)$ and 4 bits for $2C^\dagger$).
The $16$ random bits on the other hand label $2E_{8}$. The encoding can be done again
via construction $A$, since we have from (\ref{eq:e82})
that
\[
2E_8=2\sqrt{2}\mathbb{Z}^{8}+\sqrt{2}(8,4,4),
\]
for which we need $4$ random bits for $\sqrt{2}\cdot(8,4,4)$ and the rest for $4\mathbb{Z}^{8}$
(in particular, we need a minimum of 4 random bits).

\subsection{An 8-dimensional $N$-level nested lattice code construction}

In the above example, Alice could choose the number of random bits to be sent since
she knew Eve's $\SNR$.
Suppose now a scenario where Alice perfectly knows Bob's $\mathsf{SNR}$
but has no idea of Eve's $\mathsf{SNR}$, actually Alice does not even need to know that Eve is present. In this case, 
the idea we want to develop is that Alice
can decide a hierarchy of secret bits, ranking the data bits from the most secret to the least, and encode them accordingly. 
In this case, the role of the random bits in the coset coding scheme is played by the least secure bits, 
whose cardinality depends on Eve's $\SNR$. This idea has been formulated, from an information theoretic point of view, 
in \cite{shlomo-slides}. 

We now illustrate this idea by extending the
example of Subsection \ref{sub:2-level-nested-lattice}.

\begin{table}[ht]
\noindent \begin{centering}
\begin{tabular}{|c|c|}
\hline
Lattice $\Lambda$ & Code $C$\tabularnewline
\hline
\hline
$\mathbb{Z}^{8}$ & $(8,8,1)$\tabularnewline
\hline
$D_{8}$ & $(8,7,2)$\tabularnewline
\hline
$D_{4}^{2}$ & $(8,6,2)=(4,3,2)^{2}$\tabularnewline
\hline
$L_{8}$ & $(8,5,2)$\tabularnewline
\hline
$\sqrt{2}E_{8}$ & $(8,4,4)$\tabularnewline
\hline
$2L_{8}^{\star}$ & $(8,3,4)=(8,5,2)^{\bot}$\tabularnewline
\hline
$2\left(D_{4}^{\star}\right)^{2}$ & $(8,2,4)=(4,1,4)^{2}=\left((4,3,2)^{\bot}\right)^{2}$\tabularnewline
\hline
$2D_{8}^{\star}$ & $(8,1,8)=(8,7,2)^{\bot}$\tabularnewline
\hline
$2\mathbb{Z}^{8}$ & $(8,0,\infty)$\tabularnewline
\hline
\end{tabular}
\par\end{centering}
\caption{\label{tab:lat-dim-8}Construction $A$ for nested $8-$dimensional lattices}
\end{table}

First, we need a tower of nested lattices in dimension $8$ \cite{Hong-TCM}.
We give in Table \ref{tab:lat-dim-8} the construction $A$ of all nested lattices
from $\mathbb{Z}^{8}$ to $2\mathbb{Z}^{8}$. This table is read by using a
generic binary construction $A$
\begin{equation}
\Lambda=2\mathbb{Z}^{8}+C
\label{eq:construction-A}
\end{equation}
where $C$ is an $(8,k,d)$ code whose generator matrix $G_k$ can be obtained by 
taking the $k$ last rows of the following matrix $G$:
\begin{equation}\label{eq:G}
G=\left[\begin{array}{cccccccc}
0 & 0 & 0 & 0 & 0 & 0 & 0 & 1\\
0 & 0 & 0 & 1 & 0 & 0 & 0 & 1\\
0 & 0 & 0 & 0 & 0 & 0 & 1 & 1\\
0 & 0 & 0 & 0 & 0 & 1 & 0 & 1\\
0 & 0 & 1 & 1 & 0 & 0 & 1 & 1\\
0 & 1 & 0 & 1 & 0 & 1 & 0 & 1\\
0 & 0 & 0 & 0 & 1 & 1 & 1 & 1\\
1 & 1 & 1 & 1 & 1 & 1 & 1 & 1
\end{array}\right].
\end{equation}

As all codes used in Table \ref{tab:lat-dim-8} are nested codes,
all constructed lattices are nested lattices satisfying
\begin{equation}\label{eq:nested-lattices}
\mathbb{Z}^{8}\supset D_{8}\supset D_{4}^{2}\supset L_{8}\supset \sqrt{2}E_{8}\supset 2L_{8}^{\star}\supset
2\left(D_{4}^{2}\right)^{\star}\supset 2D_{8}^{\star}\supset2\mathbb{Z}^{8}.
\end{equation}
Since this nested chain is periodic ($2\mathbb{Z}^{8}$ is just a scaled version
of $\mathbb{Z}^{8}$), we can shift it in such a way that we obtain the chain
\begin{equation}
\frac{1}{\sqrt{2}}E_{8}\supset L_{8}^{\star}\supset\left(D_{4}^{2}\right)^{\star}\supset D_{8}^{\star}\supset 
\mathbb{Z}^{8}\supset D_{8}\supset D_{4}^{2}\supset L_{8}\supset \sqrt{2}E_{8}.
\label{eq:new-chain}
\end{equation}
While in Subsection \ref{sub:2-level-nested-lattice} we considered
\[
\frac{1}{\sqrt{2}}E_8 \supset \sqrt{2}E_8,
\]
we now get the same two nested (and scaled) lattices but with a finer chain of lattices in between.

To transmit $k$ information bits to Bob, that is
\[
R_s=\frac{2k}{8}
\]
bits per complex channel use, Alice chooses again $\Lambda_b=E_8$, and needs
\[
\left|\frac{1}{\sqrt{2}}E_8/\Lambda_e\right|=2^k.
\]
For instance, to send $k=1$ bit, Alice takes from (\ref{eq:new-chain}) the lattice $\Lambda_{e}=L_{8}^{\star}$.
Similarly, for $k=2$ bits, she uses $\Lambda_{e}=\left(D_{4}^{2}\right)^{\star}$.

Since Alice does not know Eve's $\SNR$, she decides the total rate $R$, based on the channel to Bob.
Suppose that Alice wants to encode a sorted block of $\ell$ information bits 
$\sv=(s_0,s_1,\ldots,s_{\ell-1})$, where by sorted we mean that the bit order matters: 
the bits are ranked in decreasing order of confidentiality, that is the first 
bit is the most confidential. 

Let us start by showing how the coding is done using $\Lambda_b=\ZZ^8$. The extension to $\Lambda_b=E_8$
will follow. 
\paragraph{Lattice coding when $\Lambda_{b}=\mathbb{Z}^8$\label{sec:Z8-encoding}}
Let us write
\[
\ell=8q+r,~0\leq r <8
\]
obtained by Euclidean division of $\ell$ by 8, and accordingly we form $q$ blocks 
$\sv_m=(s_{8m},\ldots,s_{8m+7})$ of 8 bits each, $m<q$ and get an extra block 
$\sv_q=(s_{8q},\ldots,s_{8q+r-1},0,\ldots,0)$ containing $r$ bits.
Each of the $q$ blocks of bits is encoded using the generator matrix $G$ in 
(\ref{eq:G}):
\[
\cv_m =\sv_m G,~\cv_q = \sv_q G
\]
and the final transmitted point is
\begin{equation}\label{eq:final-transmitted}
  \xv=\sum_{m=0}^{q}2^{m}\cv_{m}
\end{equation}
translated by a constant
vector, depending on the constellation, so that the mean value of
the constellation is $0$.

Let us now see how it works. 
Let $\gv_0,\gv_1,\dots,\gv_7$ denote the rows of the matrix $G$. Thus, 
\begin{equation}\label{eq:transmitted-word}
 \cv_m=s_{8m}\gv_0+s_{8m+1}\gv_1+\dots +s_{8m+7}\gv_7,~,m=0,1,\dots , q-1
\end{equation}
and similarly
\begin{equation}\label{eq:transmitted-word-end}
 \cv_q=s_{8q}\gv_0+s_{8q+1}\gv_1+\dots +s_{8q+r-1}\gv_{r-1},
\end{equation}
so that the final transmitted point of Equation (\ref{eq:final-transmitted}) can now be written as
\begin{eqnarray*}
  \xv&=&\sum_{m=0}^{q-1}2^{m}\left(s_{8m}\gv_0+s_{8m+1}\gv_1+\dots +s_{8m+7}\gv_7\right)
+2^q \left(s_{8q}\gv_0+s_{8q+1}\gv_1+\dots +s_{8q+r-1}\gv_{r-1}\right)\\
&=&s_0\gv_0+\left(s_{1}\gv_1+s_{2}\gv_2+\dots +s_{7}\gv_7+2\sum_{m=1}^q 2^{m-1} \cv_m\right)\in s_0\gv_0+D_8.
\end{eqnarray*}

Indeed, 
\begin{enumerate}
 \item The vector $\gv_0=(0,0,0,0,0,0,0,1)\notin D_8$.
 \item The term $s_{1}\gv_1+s_{2}\gv_2+\dots +s_{7}\gv_7$ is in $D_8$ since the $7\times 8$ matrix whose rows are 
$\gv_1,\dots , \gv_7$ is the generator matrix of the $(8,7,2)$ code which yields $D_8$ via construction $A$. 
\item The last term is obviously in $2\ZZ^8$ which is contained in $D_8$. 
\end{enumerate}
If $s_0=0$, $\xv\in D_8$, else $\xv\in D_8+\gv_0$ and the minimum squared Euclidean distance between $D_8$ and 
$D_8+\gv_0$ is equal to 1 which is the minimum distance of $\ZZ^8=\Lambda_b$. Consequently $s_0$ is the bit most sensitive 
to noise, that is the one with highest bit error probability. Note that, from a reliability point of view $s_0$ is the 
worst bit whereas it is the best one, from a security point of view.  

Let us repeat the process. We have 
\[
 s_{1}\gv_1+\left(s_{2}\gv_2+\dots +s_{7}\gv_7+2\sum_{m=1}^q 2^{m-1} \cv_m\right)\in s_1\gv_1+\left(D_4\right)^2
\]
and the minimum squared Euclidean distance between $\left(D_4\right)^2$ and 
$\left(D_4\right)^2+\gv_1$ is the minimum distance of $D_8$, that is $2$. The probability of correct decision on the 
bit $s_1$ is then higher than for $s_0$. 

This process is iterated for $s_j,~j=2,3,\dots,7$ where the lattice corresponding to the bit $s_j$ has a minimum squared 
Euclidean distance larger or equal to the one of the bit $s_{j-1}$. When reaching the bit $s_8$ ($m=1$), we get the lattice 
$2\ZZ^8$ (recall that we had $\ZZ^8$ for $s_0$). The chain of lattices obtained for $\sv_1$ is then the same one as for $\sv_0$
scaled by a factor of 2. More generally, the chain of lattices for $\sv_m$ is the same one as for $\sv_0$
scaled by a factor of $2^m$, that is 
\[
\Lambda_{m,\kappa}=2^{m+1}\mathbb{Z}^{8}+2^{m}\left(8,\kappa,d\right)
\]
based on the
chain of codes $(8,\kappa ,d)$ of Table \ref{tab:lat-dim-8} by using a scaled
construction $A$. 
%
%
The last block of $r$ bits will encode the cosets of the code $\left(8,(8-r),d\right)$ giving the transmitted point
\begin{equation*}
 \xv=\left(\left(\sum_{m=0}^{q-1}2^{m}\cv_{m}\right)
+2^q \left(s_{8q}\gv_0+s_{8q+1}\gv_1+\dots +s_{8q+r-2}\gv_{r-2}\right)\right)
+2^q s_{8q+r-1}\gv_{r-1}
\end{equation*}
where, as above, the bit $s_{8q+r-1}$ decides whether $\xv\in\Lambda_{q,8-r-1}$ or $\xv\in\Lambda_{q,8-r-1}+\gv_{r-1}$. 

Now, 
\begin{equation*}
 \xv=\left(\left(\sum_{m=0}^{q-1}2^{m}\cv_{m}\right)
+2^q \sum_{j=0}^{r-1}s_{8q+j}\gv_j\right)
+2^q \left(s_{8q+r}\gv_{r}+\dots+s_{8q+7}\gv_{7}\right)
\end{equation*}
where the bits $s_{8q+r},\dots,s_{8q+7}$ label points of the lattice 
\[
\Lambda_{s}=2^{q+1}\mathbb{Z}^{8}+2^q\left(8,8-r,d\right)
\] 
whereas the other bits label the cosets 
in $\Lambda_b/\Lambda_s$ whose coset representatives are chosen in the Voronoi cell of the point ${\bf 0}$ in $\Lambda_s$. 
Since $s_{8q+r}=\dots=s_{8q+7}=0$, they label the point ${\bf 0}$ in $\Lambda_s$ which can be interpreted as saying 
that $\xv$ is in the Voronoi cell of the point ${\bf 0}$ in $\Lambda_s$. 
In other words, the Voronoi cell of the point ${\bf 0}$ in $\Lambda_s$ is the shaping region $\mathcal{R}$ of 
the transmitted constellation \cite{Forney-Voronoi}.

\paragraph{Lattice coding when $\Lambda_{b}=E_{8}$}
We now extend the above encoding to the case where $\Lambda_{b}=E_{8}$. 
Take the block of information bits  $\sv=\left(s_{0},s_{1},s_{2},\ldots,s_{\ell-1}\right)$ and 
prepend $4$ bits equal to $0$ to form 
\[
\tilde{\sv}=\left(0,0,0,0,s_{0},s_{1},s_{2},\ldots,s_{\ell-1}\right).
\]
As above, we first compute 
\[
 \cv_0=\left(0,0,0,0,s_{0},s_{1},s_{2},s_{3}\right) G, 
\]
a codeword of the Reed-Müller code $(8,4,4)$, which, by construction $A$ gives a lattice point in $E_8$, 
after which the whole encoding procedure described for $\ZZ^8$ holds.


%
%

\section{Conclusion}
In this paper, we considered coding strategies based on lattices, for the Gaussian wiretap channel. 
From the expression of the eavesdropper probability of correct decision, we derived the 
so-called secrecy gain, a new lattice invariant related to theta series, which characterizes the amount of confusion that 
lattice coding introduces at the eavesdropper. Since theta series of even unimodular lattices are well-understood, 
we focused, in this paper, on the study of the secrecy gain of even unimodular lattices: we provided explicit 
examples and an asymptotic analysis which shows that the secrecy gain grows exponentially in the lattice dimension. 
Finally, worked out coding examples were given.  

%
%

\section*{Acknowledgment}

Part of this work was done while J.-C. Belfiore and P. Sol\'e were visiting 
the Division of Mathematical Sciences at the Nanyang Technological University, 
Singapore.

The authors would like to thank S. Robins for fruitful discussions. 

The research of F. Oggier is supported in part by the Singapore National
Research Foundation under Research Grant NRF-RF2009-07 and NRF-CRP2-2007-03,
and in part by the Nanyang Technological University under Research
Grant M58110049 and M58110070.

%
%

\appendix

In this appendix, we review results that are needed to manipulate sums of periodic 
functions over lattices. In particular, we will detail the Poisson summation formula over lattices 
and the Jacobi formula. 

Consider the function $F(\xv)=\sum_{\mv\in\ZZ^n}f(\mv+\xv)$ which is periodic
over $[0,1]^n$, for $f$ a well-behaved function, that is satisfying \cite{Ebeling}
\begin{enumerate}
 \item $\int_{\RR^n} \left| f(x) \right| dx < \infty $
 \item $\sum_{\mv\in\ZZ^n} \left| f(\mv+{\bf u}) \right| $ converges uniformly for all ${\bf u}$ belonging
to a compact subset of $\RR^n$.
\end{enumerate}%
It has a Fourier series
$F(\xv)=\sum_{\nv\in\ZZ^n}a_\nv e^{2\pi i \langle\nv,\xv\rangle}$, where
\begin{eqnarray*}
a_\nv &=& \int_{[0,1]^n}e^{-2\pi i \langle\nv,\yv\rangle}F(\yv)d\yv\\
      &=& \sum _{\mv\in\ZZ^n}\int_{[0,1]^n}e^{-2\pi i \langle\nv,\yv\rangle}
           f(\mv+\yv)d\yv\\
      &=&\sum _{\mv\in\ZZ^n}\int_{[0,1]^n+\mv}e^{-2\pi i \langle\nv,\uv-\mv\rangle}
           f(\uv)d\uv \\
      &=& \int_{\RR^n}e^{-2\pi i \langle\nv,\uv\rangle}f(\uv)d\uv =\hat{f}(\nv)
\end{eqnarray*}
where $\hat{f}(\nv)$ is the Fourier transform of $f$, which is such that we can
invert the sum and the integral in the second step, and reconstruct the integral
in the fourth step. Thus
\[
F(\xv)=\sum_{\mv\in\ZZ^n}f(\mv+\xv)=
       \sum_{\nv\in\ZZ^n}a_\nv e^{2\pi i \langle\nv,\xv\rangle}=
       \sum_{\nv\in\ZZ^n}\hat{f}(\nv) e^{2\pi i \langle\nv,\xv\rangle}
\]
which yields, in $\xv={\bf 0}$, the so-called {\em Poisson summation formula}:
\begin{equation}\label{eq:poisson}
\boxed{
\sum_{\mv\in\ZZ^n}f(\mv)=\sum_{\nv\in\ZZ^n}\hat{f}(\nv).
}
\end{equation}

One can be more general and consider summing $f$ on the points of an arbitrary
lattice $\Lambda$, say with generator matrix $M$, instead of $\ZZ^n$:
\[
\sum_{\mv\in\Lambda}f(\mv)=\sum_{\xv\in\ZZ^n}f(M\xv)=
\sum_{\yv\in\ZZ^n}\widehat{f\circ M}(\yv)
\]
using (\ref{eq:poisson}). Now
\begin{eqnarray*}
\widehat{f\circ M}(\yv)
&=&\int_{\RR^n}e^{-2\pi i \langle\yv,\xv\rangle}f(M\xv) d\xv\\
&=&|\det(M)|^{-1}\int_{\RR^n}e^{-2\pi i \langle (M^{-1})^T\yv,\uv\rangle}f(\uv)d\uv\\
&=&|\det(M)|^{-1}\hat{f}((M^{-1})^T\yv),
\end{eqnarray*}
giving the {\em Poisson summation formula for lattices}:
\begin{equation}\label{eq:poisslatt}
\boxed{
\sum_{\mv\in\Lambda}f(\mv)=|\det(M)|^{-1}\sum_{\nv\in\Lambda^\star}\hat{f}(\nv)
}
\end{equation}
where $\Lambda^\star$ has generator matrix $(M^{-1})^T$. 
The lattice $\Lambda^\star$ is the dual lattice of $\Lambda$ (see also Definition~\ref{def:dual}).

Let $\Theta_\Lambda(y)=\sum_{\rv\in\Lambda}e^{-\pi y ||\rv||^2}$ be the theta
series of $\Lambda$ with generator matrix $M$, which we rewrite as
$\Theta_\Lambda(y)=\sum_{\rv\in\Lambda}f(\rv)$, so as to apply (\ref{eq:poisslatt}):
\[
\Theta_\Lambda(y)= |\det(M)|^{-1}\sum_{\nv\in\Lambda^\star}\hat{f}(\nv)
\]
where
\begin{eqnarray*}
\hat{f}(\nv)&=&\int_{\RR^n}e^{-2\pi i \langle\nv,\xv\rangle}f(\xv)d\xv\\
         &=&\int_{\RR^n}e^{-2\pi i \langle\nv,\xv\rangle}e^{-\pi y ||\xv||^2}d\xv\\
         &=&\prod_{j=1}^n\int_{\RR}e^{-2 i \pi n_jx_j-\pi y x_j^2}dx_j\\
         &=& \left(\frac{1}{\sqrt{y}}\right)^ne^{-\pi y || \nv||^2/y}.
\end{eqnarray*}
We conclude that
\[
\Theta_\Lambda(y)= |\det(M)|^{-1}\sum_{\nv\in\Lambda^\star}\left(\frac{1}{\sqrt{y}}\right)^ne^{-\pi y || \nv||^2/y},
\]
which yields the {\em Jacobi's formula} \cite{CS-98}
\begin{equation}\label{eq:jacobi}
\boxed{
\Theta_\Lambda(y)=|\det(M)|^{-1}\left(\frac{1}{\sqrt{y}}\right)^n
\Theta_{\Lambda^\star}(1/y),}
\end{equation}
connecting the theta series of a lattice and its dual.

%
%

\end{document}